\documentclass[sigconf]{acmart}

\usepackage[utf8]{inputenc}
\usepackage{caption}
\usepackage{subcaption}
\usepackage{appendix}

\usepackage[resetlabels]{multibib}
\newcites{game}{Ludography}
\nocitegame{*}
\newcommand{\citegameprefix}{G}
\usepackage{xparse}
\let\origcitegame\citegame
\RenewDocumentCommand{\citegame}{O{} O{} m}{%
  \renewcommand{\citenumfont}[1]{\citegameprefix##1}%
  \origcitegame[#1][#2]{#3}%
  \renewcommand{\citenumfont}[1]{##1}%
}
\usepackage{ragged2e}
\usepackage{soul}

\usepackage{placeins}  
\usepackage{floatflt}
\usepackage{enumitem}
\usepackage{booktabs}
\usepackage{tabularx}
\AtBeginDocument{%
  }

\copyrightyear{2026}
\acmYear{2026}
\setcopyright{cc}
\setcctype{by}
\acmConference[CHIWORK '26]{Proceedings of the 5th Annual Symposium on Human-Computer Interaction for Work}{June 22--25, 2026}{Linz, Austria}
\acmBooktitle{Proceedings of the 5th Annual Symposium on Human-Computer Interaction for Work (CHIWORK '26), June 22--25, 2026, Linz, Austria}
\acmDOI{10.1145/3808045.3808068}
\acmISBN{979-8-4007-2429-9/2026/06}

\begin{document}


\title{Upskilling with Generative AI: Practices and Challenges for Freelance Knowledge Workers}


\author{Kashif Imteyaz}
\email{imteyaz.k@northeastern.edu}
\affiliation{%
  \institution{Civic AI Lab, Northeastern University}
  \city{Boston}
  \state{Massachusetts}
  \country{USA}
}

\author{Isabel Lopez}
\email{isabellhur@gmail.com}
\affiliation{%
  \institution{CICESE}
  \city{Ensenada}
  \state{Baja California}
  \country{Mexico}
}

\author{Nakul Rajpal}
\email{rajpal.n@northeastern.edu}
\affiliation{%
  \institution{Civic A.I. Lab, Northeastern University}
  \city{Boston}
  \state{Massachusetts}
  \country{USA}
}

\author{Hunjun Shin}
\email{shin.hu@northeastern.edu}
\affiliation{%
  \institution{Civic A.I. Lab, Northeastern University}
  \city{Boston}
  \state{Massachusetts}
  \country{USA}
}

\author{Saiph Savage}
\email{s.savage@northeastern.edu}
\affiliation{%
  \institution{Civic A.I. Lab, Northeastern University}
  \city{Boston}
  \state{Massachusetts}
  \country{USA}
}
\affiliation{%
  \institution{Universidad Nacional Autonoma de Mexico (UNAM)}
  \city{Mexico City}
  \country{Mexico}
}
\renewcommand{\shortauthors}{Imteyaz et al.}

\begin{abstract}

Freelance workers must continually acquire new skills to remain competitive in online labor markets, yet they lack the organizational training, mentorship, and infrastructure available to traditional employees. Generative AI-powered tools like ChatGPT are reshaping market skill demands while also offering new forms of on-demand learning support to meet those demands. Despite growing interest in AI-powered learning tools, little is known about how freelancers actually use these tools to learn, the challenges they encounter, and how generative AI for learning interacts with precarity and competition in platform-based work. We present a mixed-methods study combining a survey and semi-structured interviews with freelance knowledge workers. Grounded in self-directed learning theory, we examine how freelancers integrate generative AI tools into their learning practices. Our findings show that freelancers increasingly rely on generative AI to structure learning and support exploratory skill acquisition, but do not treat it as their primary learning resource due to inconsistency, lack of contextual relevance, and verification overhead. We identify a shift from \textit{learning as growth to learning as survival}, where upskilling is oriented toward immediate market viability rather than long-term development. We also surface a structural challenge we term \textit{invisible competencies}, in which workers acquire skills through generative AI tools but lack credible ways to signal or validate these skills in competitive freelance markets. Based on these insights, we offer design recommendations for generative AI-powered learning tools for freelancers.

\end{abstract}

\begin{teaserfigure}
  \includegraphics[width=\textwidth]{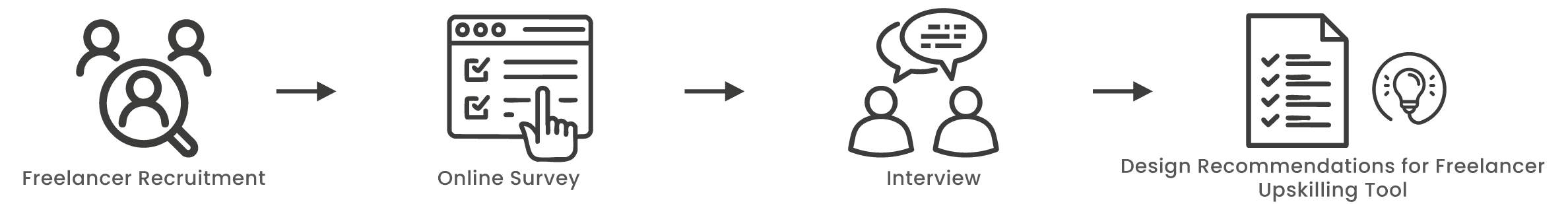}
  \Description{Flowchart showing four-stage methodology from left to right: (1) Freelancer Recruitment with people and magnifying glass icon, (2) Online Survey with checklist form, (3) Interview with two people and speech bubbles, and (4) Design Recommendations for freelancer upskilling tool shown as document and lightbulb. Arrows connect each stage sequentially.}
  \caption{Overview of our exploratory mixed method study}
  \label{fig:teaser}
\end{teaserfigure}





\maketitle

\section{Introduction}
Freelance knowledge workers face persistent challenges in maintaining the ongoing skill development required to remain competitive in online labor markets \cite{blaising2021making}. Unlike traditional corporate employees, who often benefit from institutional support such as structured AI literacy workshops, supervised tool onboarding, peer learning cohorts, or mentorship aligned with organizational goals \cite{Campana2014, wilkins2022gigified, zhang2025knowledge}, freelancers must independently identify, evaluate, and acquire new skills to sustain their livelihoods \cite{chiang2018crowd, Uncertainty10.1145/3290607.3312922}. Without access to organizational infrastructure or formal guidance \cite{roy2020future, wilkins2022gigified}, freelancers bear full responsibility for managing their own learning trajectories \cite{blaising2021making, alvarez2023understanding}. This burden is intensified by the competitive nature of freelance platforms \cite{roy2020future, emotionalSupport}, where sharing knowledge about effective upskilling strategies may directly undermine one’s own economic opportunities \cite{pleskac2021ecology, wood2019good}. As a result, peer learning practices that are common in traditional workplaces become difficult to sustain when workers are competing for the same contracts \cite{munoz2022platform}.

These long-standing challenges have become more urgent with the rapid advancement of Large Language Models (LLMs) such as ChatGPT and other AI-powered learning tools. LLMs are reshaping knowledge work across domains \cite{peng2023impactaideveloperproductivity, tamkin2021understanding, brynjolfsson2025generative}, introducing new forms of work while simultaneously redefining the skills required to perform existing tasks \cite{ Gruenewald2025Reskilling, zhang2025knowledge, shen2026ai}. As these systems become increasingly integrated in everyday knowledge-based labor \cite{prasad2024towards, Claudepaper}, freelancers must not only learn how to use new AI tools, but also develop novel competencies such as prompt design, AI-assisted content production, and the integration of LLM outputs into client-facing deliverables \cite{popiel2017boundaryless, knight2024impact, shen2026ai}. These pressures are further amplified by the continued expansion of the freelance economy \cite{burke2020relationship, upwork_freelancing_2023}. By 2025, more than 1.56 billion people globally engaged in freelance or independent work, reflecting a large-scale shift toward contingent and flexible labor arrangements \cite{statista_freelancers, upwork_freelancing_2023}. In this context, the ability to learn and adapt quickly is no longer optional but central to freelancers’ economic survival.

At the same time, LLMs present a paradox: while they increase uncertainty by accelerating skill obsolescence and intensifying competition \cite{autor2015there,susskind2020world,acemoglu2022artificial}, they also create new opportunities for personalized, on-demand learning support \cite{kasneci2023chatgpt, LLM-BasedPythonTutorCHI25}. However, most existing generative AI learning systems are designed for traditional organizational settings \cite{Morandini2023}, assuming stable employment structures, managerial guidance, and collective learning environments \cite{blau2008relation,eraut2004informal}. As a result, little is known about how LLMs are reshaping learning practices for freelance workers who operate outside these institutional contexts. We lack empirical understanding of how freelancers use LLMs to learn, what challenges they encounter, and how AI-mediated learning intersects with precarity, competition, or self-directed work. Understanding these dynamics matters, as learning infrastructures that fail to account for freelance realities risk widening inequalities in access to skills, income stability, and long-term career mobility. Studying how LLMs are transforming the learning landscape for freelance workers is therefore essential for designing AI-assisted learning systems that genuinely support independent workers rather than exclude them. 

While freelancers have always faced challenges in self-directed learning \cite{Uncertainty10.1145/3290607.3312922, blaising2021making}, AI-mediated learning introduces new dynamics that prior work has not examined \cite{ChallangeinLLMforlearning}. First, the learning resource itself is generative and opaque \cite{zamfirescu2023johnny}, producing content without provenance or editorial review \cite{ChallangeinLLMforlearning}, which shifts the learner's burden from finding resources to verifying them \cite{VerificationofLLMLabels, 10.1145/3613905.3636293}. Second, skills acquired through AI-mediated learning leave no social trace and attribution~\cite{wilkins2022gigified, he2025contributions, polimetla2026paradigm}, making them uniquely difficult to signal in markets that rely on credentials and peer endorsement~\cite{munoz2022platform, dolata2025moreAttention}. Our contribution lies in identifying what changes when freelancers' learning is mediated by generative AI systems with these characteristics.

To address this knowledge gap, we conducted a mixed-methods study combining survey data (n=71) and follow-up interviews (n=20) with platform-based knowledge workers, professionals who perform information-based tasks such as writing, data analysis, design, and software development. We ground our analysis in self-directed learning theory \cite{garrison1997self}, which is particularly well-suited for understanding how freelancers adapt to rapidly evolving landscapes. Self-directed learning theory conceptualizes learning as a process where individuals take primary responsibility for planning, conducting, and evaluating their learning experiences \cite{garrison1997self}. Since freelancers typically lack access to formal training, mentorship, and support systems available to traditional employees \cite{blaising2022managing}, they must take full responsibility for their learning by setting goals, identifying resources, and evaluating progress. Self-directed learning's emphasis on real-world problem-solving and practical application provides a useful lens for examining how freelancers navigate constant upskilling pressures in competitive and isolating environments. Fig.~\ref{fig:teaser} presents an overview of our study. Our research focuses on the following research questions:

\begin{enumerate}
    \item How do freelance knowledge workers approach skill development in work environments increasingly shaped by generative AI adoption?
    \item What barriers do freelancers encounter when trying to access learning opportunities for skill development in AI-driven work environments?
    \item How do freelancers envision future learning tools that can support their skill development in the context of widespread generative AI adoption?
\end{enumerate}

In our mixed-methods study, we found that freelancers increasingly use generative AI tools to support skill development. In particular, they rely on these tools to structure learning plans that break skills into manageable steps and fit learning into demanding work schedules. We also observed a shift toward \emph{learning-as-survival}, where upskilling is oriented toward immediate market viability rather than long-term professional growth. 

At the same time, freelancers did not treat generative AI as their primary or fully trusted learning source. A key limitation they felt with generative AI tools was the lack of contextual personalization. LLM-generated learning content often felt generic and did not reflect freelancers’ specific, situated needs \cite{haraway2013situated}. To fill this gap, freelancers turned to peer learning for more relevant, experience-based knowledge. However, peer learning was constrained by the competitive nature of freelance work, which made many freelancers hesitant to share what they know \cite{asker2016competitive}. Within this landscape, freelancers increasingly described generative AI as a lower-risk option for exploratory learning, enabling them to learn and test skills privately while relying on peers for validation, access to opportunities, and legitimacy-building. 

We also found that AI-based learning introduced additional burdens for freelancers. While prior work has noted that generative AI can introduce new costs alongside productivity gains \cite{simkute2025ironies}, we find that these costs are amplified and reconfigured in freelance learning contexts. Freelancers reported information overload, inconsistent guidance, and the need to verify AI-generated content, increasing the effort required to learn effectively. Beyond these verification demands, freelancers described a structural challenge that we call \emph{invisible competencies}. Although freelancers can acquire real skills through AI-assisted learning, these capabilities are often difficult to signal or validate in freelance labor markets. Concerns about reputation, along with the risks of visible experimentation, further discourage open learning \cite{mohlmann2021algorithmic, jarrahi2021algorithmic, SkillDevelopNecessity}.

In response, freelancers envisioned generative AI–based learning tools that reduce verification burden, support strategic learning under market pressure, and help translate AI-acquired skills into clear and credible signals of competence for clients. Based on our findings, we provide design recommendations for building generative AI-based learning tools that better meet the needs of freelancers. 

In this paper, we contribute:
\begin{itemize}
    \item A mixed-methods study of freelancers’ upskilling practices, examining how generative AI is used as a learning resource in competitive, platform-mediated work contexts.
    \item A conceptual framing of freelancer upskilling that characterizes a shift from learning-as-growth toward \emph{learning-as-survival}, and identify \textit{invisible competencies} as a key consequence of AI-assisted learning, where genuinely acquired skills remain difficult to demonstrate or validate in freelance labor markets.
    \item Design recommendations for future generative AI tools that better support freelancers' learning needs, grounded in freelancers' own experiences and aspirations.
\end{itemize}

\section {Related Work}
Our research focuses on the algorithmic management of platform labor, AI-driven job displacement, self-directed learning theory, and the use of AI for upskilling. We review each area to situate our study on how freelancers navigate upskilling.

\subsection{Algorithmic Management and Platform Labor}
Platform-mediated freelance knowledge work is structured by algorithmic systems that shape workers’ visibility, evaluation, and access to economic opportunities \cite{mohlmann2021algorithmic, wood2019good}. Research shows that algorithmic management allows for managerial control through ranking systems, automated task allocation, and data-driven oversight, which alters everyday decision making \cite{WorkingWithMachines, mohlmann2021algorithmic}. Prior research on platform-based work shows that workers experience workplace surveillance, information asymmetries, and opaque algorithmic processes under asymmetric power dynamics \cite{do2024designing, zhang2022algorithmic}. Within these systems, reputation metrics and client reviews function as core mechanisms of algorithmic management \cite{jarrahi2021algorithmic, wood2019good}. Freelancers place substantial economic value on maintaining strong ratings \cite{reviewsAM, cotter2020algorithmic}, which can constrain upskilling by discouraging them from taking on projects that require skills they are still developing. Platform work is also marked by broader forms of precarity, including emotional labor and the absence of institutional support \cite{BrushItOff, munoz2022new}. 

Recent work shows that freelance workers seek ways to reclaim their agency within these systems \cite{zhang2024data, imteyaz2026co}, from designing worker-controlled data institutions to articulating principles for more transparent and worker-centered algorithmic management \cite{YouApp, zhang2022algorithmic, hsieh2023co}, and calling for institutional and policy-level reforms to address algorithmic opacity and power asymmetries \cite{nagaraj2025rideshare}. This body of work positions platform labor as a sociotechnical environment that directly shapes learning: algorithmic instability discourages long-term skill investment, reputational risk makes freelancers reluctant to experiment with unfamiliar skills in client-facing work, and limited structural support means they must navigate these decisions alone \cite{zhang2022algorithmic, wood2019good, wood2018workers, calacci2022organizing, kim2026occupational}.

\subsection{AI Automation and Job Displacement}

Understanding AI automation and job displacement has become increasingly critical as Large Language Models (LLMs) and agentic tools fundamentally transform knowledge work \cite{Claudepaper, anthropiccowork, brynjolfsson2025generative, pachera2026co}. Unlike previous automation waves that primarily affected manual labor \cite{autor2015there}, current LLM-based tools such as ChatGPT and Claude demonstrate unprecedented capabilities in automating entry-level cognitive tasks and routine knowledge work \cite{shao2025future}, roles that traditionally served as gateways to professional careers \cite{woodruff2024knowledge, wagman2025generative}. This shift introduces immediate competitive pressures, particularly for vulnerable worker populations lacking institutional support \cite{do2024designing, sumbal2024wind, wood2018workers, flores2025impact, imteyaz2024human}. Platform-mediated labor markets show measurable disruption within months of new LLM releases. On Upwork, writing-related freelancers experienced a 2\% drop in monthly jobs and a 5.2\% drop in monthly earnings following ChatGPT's release, with image-related freelancers seeing even steeper declines after DALL-E 2 and Midjourney~\cite{hui2024short-term}. Recent HCI research highlights the complex dynamics of augmentation, automation, and adaptation beyond simplistic replacement narratives~\cite{yun2025generative, kyi2025governance, woodruff2024knowledge, imteyaz2024human}.

However, existing studies largely focus on macroeconomic trends and platform analytics \cite{jobDisplacement2024, teutloff2025winners, brynjolfsson2025generative}, offering limited insight into how individual freelance workers respond to these disruptions. From an HCI perspective, this gap underscores the need to understand how freelancers engage in situated strategies to remain viable in an AI-disrupted labor market \cite{shao2025future, yun2025generative}. Specifically, little is known about how freelancers adapt to AI-driven disruptions or develop strategies to upskill and remain competitive.

\subsection{Self-Directed Learning in Freelance Context}
Self-directed learning (SDL) theory provides a framework for understanding how freelancers navigate skill development without institutional support. Garrison \cite{garrison1997self} defines SDL as a process where learners take primary responsibility for planning, conducting, and evaluating their learning experiences. The framework comprises three interrelated components: \textit{self-management} (structuring and organizing learning activities), \textit{self-monitoring} (evaluating and reflecting on progress), and \textit{motivation} (sustaining effort and goal orientation). This framework emphasizes learner autonomy in setting goals, selecting resources, and monitoring progress, characteristics that align with the independent nature of freelance work \cite{roy2020future}. SDL has been recognized as crucial for preventing skill obsolescence and supporting upskilling in volatile economic conditions \cite{morris2019}.

However, freelancers' self-directed learning occurs under structural constraints \cite{gussek2024freelancer}. Unlike traditional employees who access skill development through formal training, peer learning cohorts, curated resources, and protected experimentation time \cite{lang2023workforce, reskilling_upskilling_2022}, freelancers must independently identify skill gaps, fund their upskilling, and integrate learning into immediate income-generating work \cite{chiang2018crowd}. Research identifies four key overheads that shape freelancer learning \cite{blaising2021making}: \textit{financial overhead} from income volatility; \textit{emotional overhead} from self-management stress and isolation \cite{do2024designing}; \textit{relational overhead} from limited mentorship \cite{Uncertainty10.1145/3290607.3312922, wood2019good}; and \textit{reputational overhead} from continuous impression management, where single negative reviews carry long-term consequences \cite{zhang2022algorithmic, wood2019good}. Critically, freelancers lack the \textit{social validation} that organizational contexts provide: the ability to observe peers, ask questions, and develop shared norms about appropriate AI use \cite{10.1145/3596671.3597655, FixedTeamwork, wilkins2022gigified}.

In the absence of institutional support, peer learning becomes a potentially valuable resource for freelancers \cite{chiang2018crowd}. Research on peer learning typically assumes collaborative environments where knowledge sharing is mutually beneficial \cite{qiu2025self, Trust2016}. However, competitive freelance markets complicate these assumptions. Platform dynamics, including algorithmic visibility, public profiles, and bidding systems, position freelancers as direct competitors, incentivizing knowledge hiding to preserve competitive advantage \cite{gagne2022understanding, asker2016competitive, mohlmann2021algorithmic}. This makes visible experimentation costly and encourages selective reciprocity over open sharing \cite{fan2020crowdco}.
Despite these barriers, peer learning offers something that other learning resources may not: \textit{situated knowledge} grounded in identifiable individuals with known experience and professional trajectories \cite{haraway2013situated, huang2024design}. Situated knowledge research emphasizes that knowledge claims are accountable to the social and professional conditions under which they are produced \cite{haraway2013situated}. For freelancers navigating uncertain information landscapes, knowing \textit{who} produced knowledge, and under what conditions, may be as important as the knowledge itself. Yet systematic understanding of how freelancers balance peer learning benefits against competitive risks \cite{gussek2024freelancer, systematic_gig_economy_2024, CooperationRied}, and how this intersects with emerging AI learning tools, remains limited.

\subsection{AI for Upskilling: Opportunities and Tensions}
While LLMs and AI agents introduce job displacement pressures \cite{wagman2025generative, simkute2025ironies}, they simultaneously offer unprecedented opportunities for personalized, accessible learning \cite{wang2025learnmate, LLM-BasedPythonTutorCHI25}. In HCI and learning sciences, researchers have long explored how intelligent systems can support skill development through adaptive tutoring \cite{anderson1995cognitive}, intelligent feedback \cite{kochmar2022automated}, and personalized learning paths \cite{brusilovsky2007user}. These approaches build on scaffolding theory \cite{verenikina2008scaffolding}, temporary support structures that enable learners to achieve beyond their independent capabilities \cite{wood1976role, quintana2004ideakeeper}. Recent advances in LLMs have expanded these possibilities, enabling on-demand explanations, code generation for learning programming \cite{kazemitabaar2023studying, ScaffoldingAlgorithmicusingLLM}, and conversational learning partners \cite{lytvyn2025human, wang2025learnmate, LLMCompanionsCHI24}. However, applying scaffolding principles to freelance learning reveals unique challenges. The concept of "fading" support as competence develops becomes complex when errors directly impact earnings and platform ratings \cite{zhang2022algorithmic}. LLMs often fail to balance what \citet{reiser2018scaffolding} terms structure versus problematization: they either over-scaffold through complete task completion (reducing learning opportunities) or under-scaffold through generic advice (failing to address specific skill gaps). Additionally, freelancers face challenges in evaluating LLM generated content quality, particularly when AI systems produce plausible-sounding but incorrect information \cite{VerificationofLLMLabels, danry2025deceptive}. This "verification burden" is especially acute for freelancers who must spend time validating learning materials \cite{chiang2018crowd, toxtli2021quantifying}. 

While LLMs can provide personalized instruction, it cannot replicate the motivational and social benefits of human peer learning \cite{riedl2025potential, CooperationRied}. Despite growing research on LLM-based learning tools \cite{ChallangeinLLMforlearning}, we lack an integrated understanding of how to design AI learning systems that address the unique constraints and competitive pressures that freelancers face while engaging in upskilling practices. 

\subsection{Research Gap}
While prior work offers valuable insights across these domains, the literature remains largely separate. Research on platform labor documents how algorithmic systems shape freelancer behavior but rarely examines learning practices. Studies of AI-driven displacement identify macroeconomic trends but offer little insight into how individuals actually adapt. Self-Directed Learning theory offers a useful framework for autonomous learning, though it has not been applied to contexts where learners face simultaneous competitive pressure, reputational risk, and rapidly evolving tools. HCI research has produced promising AI-powered learning systems grounded in scaffolding theory, but these have primarily targeted institutional contexts with built-in social support. Our study examines what happens in the gap between these literatures: how platform-based freelancers who lack institutional infrastructure, compete directly with peers, and face algorithmic reputation pressures, use generative AI tools for self-directed upskilling.

\section{Method}
Our IRB-approved study employed a sequential mixed-methods design \cite{creswell2017, ivankova2006using, adnin2025examining}. Initially, data were collected through an online survey, which contained both open-ended and closed-ended responses. Following this, we conducted interviews with subset of survey respondents. This approach allowed us to first identify broad patterns in the freelancer's upskilling approach, then explore the underlying mechanisms and contexts through in-depth interviews.

\subsection{Participants and Recruitment}
A total of 71 individuals participated in the survey. The majority of respondents were between the ages of 25–34 (55\%), with an almost equal distribution of gender: 52.1\% identified as female and 46.5\% as male, while one participant chose not to disclose their gender. Most participants held at least a bachelor's degree, with 60.6\% reporting this as their highest level of education. The respondents varied widely in terms of professional background, freelancing experience, and hours dedicated to freelance work (see Table \ref{tab:demographics} in Appendix). Freelance roles were diverse, with the most common areas being content creation and writing (n = 16), followed by data analysis (n = 13), consulting (n = 9), digital marketing (n = 6), Design/Creative Work (n = 5), and software development (n = 5). Additional roles included Virtual assistance/administrative support, customer support/AI automation scriptwriting, and product management. We recruited participants for our study by creating a job posting on Upwork, a prominent freelance platform \cite{blaising2021making}. Eligibility criteria required participants to: (1) have at least one year of active experience on a freelancing platform, (2) have used Generative AI tools for upskilling, and (3) be proficient in English. Additionally, participants must have engaged in upskilling practices within the last six months to advance their freelancing careers. We define upskilling as deliberate efforts to acquire new skills or deepen existing ones for professional purposes, such as taking online courses, using AI tools to learn techniques, engaging in peer knowledge exchange, obtaining certifications, or self-directed practice with new tools or technologies~\cite{beier2025workplacelearning}. These criteria ensured participants had both longitudinal platform experience and recent upskilling insights. Our survey sample reflects the practical constraints of recruiting from a freelance platform with specific eligibility requirements. Consistent with our sequential mixed-methods design \cite{creswell2017}, the survey was not intended to produce population-level generalizations but to identify broad patterns in freelancers' upskilling practices that would inform the subsequent interview phase.

We invited 20 survey respondents to participate in interviews using purposive sampling to better understand the reasons behind the patterns observed in the survey data \cite{blandford2016qualitative, campbell2020purposive}. Three findings specifically shaped our interview focus and participant selection: (1) the gap between high AI tool adoption and low preference for AI-based learning, (2) the tension between frequent peer learning practice and low preference ratings, and (3) participants' open-ended descriptions of AI-related challenges that needed deeper exploration. We selected interview participants by first stratifying the survey pool along three dimensions: freelance domain, years of freelancing experience (1-2 years to 5+ years), and AI tool usage intensity (low, moderate, high). We prioritized ensuring coverage across all freelance domains represented in the survey, then within each domain, selected participants who varied in experience level and AI usage. When multiple candidates met the same criteria, we prioritized those whose open-ended survey responses suggested they could provide richer accounts of the tensions identified in the survey data (see Table \ref{tab:participant-demographics} in the appendix). 

\subsection{Data Collection}
We developed and distributed an online survey using Google Forms. The survey included a mix of question types: multiple-choice, Likert scale, checkbox selection, and open-ended responses. Closed-ended response options were developed through a two-step process. First, we compiled survey items from prior research on freelancer skill development \cite{blaising2021making, chiang2018crowd} and platform labor challenges \cite{wood2019good}. For features valued in AI learning tools, items were informed by both prior work and established learning frameworks, including self-directed learning \cite{garrison1997self}, experiential learning \cite{kolb2014experiential}, and principles of adult learning \cite{knowles1980modern}. For example, personalized learning paths reflect self-directed learning's emphasis on learner autonomy, while project-based practice exercises align with experiential learning principles. We supplemented each item set with an "Other" option and an open-text field, allowing participants to identify features or barriers not captured in the predefined list.

The survey was divided into six sections: Demographics, Professional Background, Current Upskilling Practices, Peer Learning Challenges and Benefits, Learning Preferences and Needs, and Current Use and Opportunities of AI. We conducted semi-structured interviews guided by a protocol aligned with the survey items but designed to elicit deeper insights. On average, each session lasted approximately 60 minutes. All interviews were conducted online via Zoom and recorded for analysis. Interviews both confirmed survey-identified patterns and surfaced additional needs beyond the predefined options.

\subsection {Analysis}
To analyze the quantitative data collected from the survey, we followed standard statistical analysis techniques and reported the frequency of responses in different categories \cite{kitchenham2008personal}. Closed-ended questions related to current upskilling practices, learning preferences, and patterns of use of AI tools were analyzed using descriptive statistics in RStudio.   

For our qualitative data, which included interviews and open-ended responses, we followed the practices of reflexive thematic analysis \cite{braun2019reflecting}, conducting a bottom-up analysis of our interview transcripts and open-ended survey responses. Two researchers independently open-coded separate halves of the transcripts, focusing on our research questions regarding freelancers' upskilling practices, meeting bi-weekly to discuss emerging patterns, refine the codebook, and resolve disagreements through consensus. Consistent with Braun and Clarke \cite{braun2019reflecting}, we did not compute inter-rater reliability, focusing instead on interpretive depth and reflexive engagement. After open coding, we iteratively grouped codes into higher-level themes through axial coding, examining relationships between codes and their relevance to freelancers' upskilling processes. We continued analysis until thematic saturation was reached \cite{lowe2018quantifying}, with no substantially new insights emerging in the final interviews and themes well supported across the data. This process produced 19 axial codes, which we synthesized into 6 core themes that structure our findings. Throughout, we maintained reflexive memos documenting analytical decisions. Participant quotes are anonymized as P followed by participant ID. Our initial coding was bottom-up and data-driven. SDL entered the process during later stages of theme development, when we used Garrison's three components (self-management, self-monitoring, and motivation) as interpretive lenses to contextualize emergent themes and structure the discussion \cite{garrison1997self}.

\section{Survey Results}
We present both closed-ended and open-ended survey results that describe freelancers' perceived impact of AI on their work, patterns of upskilling, and AI-based learning preferences.

\subsection{Impact of AI in Freelance Work}

\begin{figure}[!htbp]
  \centering
  \includegraphics[width=\columnwidth,
    alt={Stacked horizontal bar. Rating 5 occupies 57 percent, rating 4 is 23 percent, rating 3 is 13 percent, rating 2 is 7 percent. No segment for rating 1.}]
    {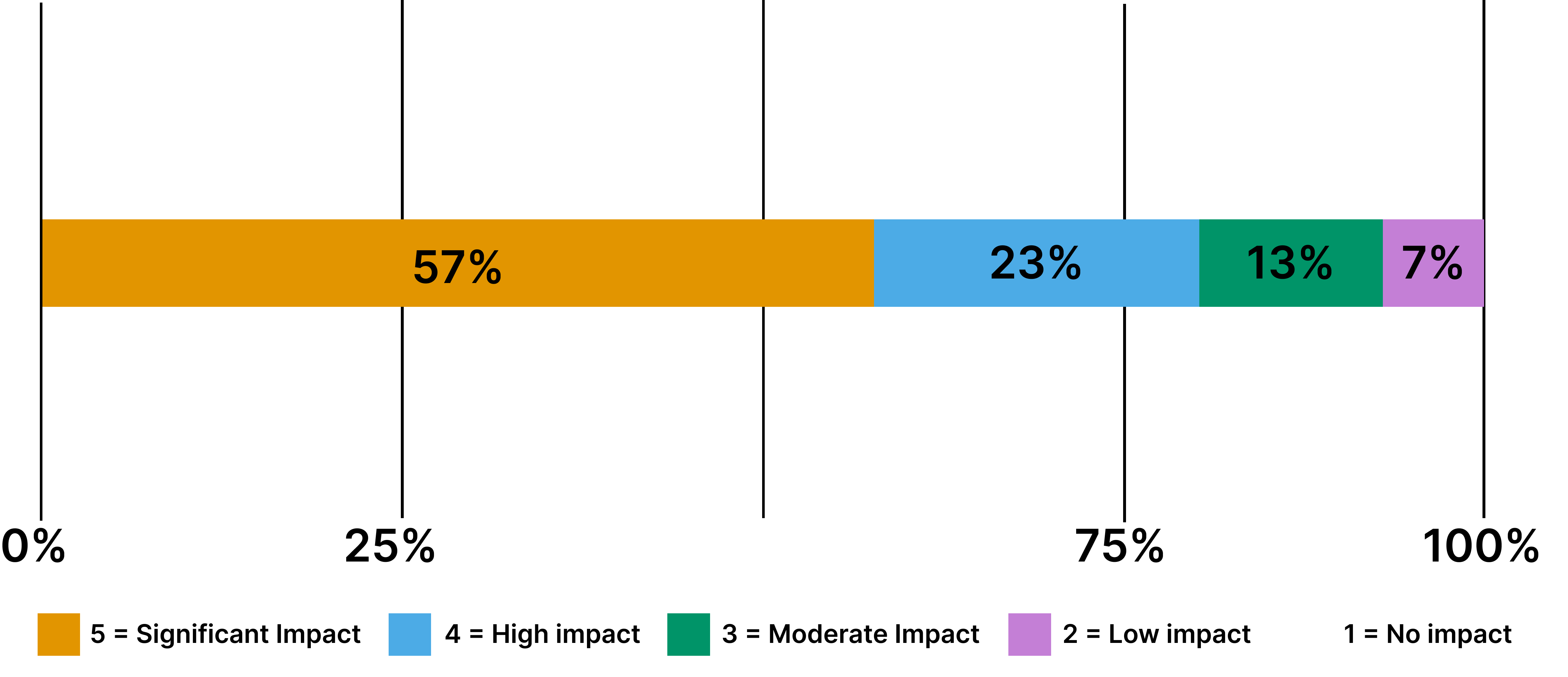}
  \caption{Perceived impact of AI on freelance work over the past year, rated from 5 (significant impact) to 1 (no impact).}
  \Description{Stacked bar showing impact ratings: 5 (significant) 57 percent, 4 (high) 23 percent, 3 (moderate) 13 percent, 2 (low) 7 percent, 1 (none) 0 percent.}
  \label{fig:AIfreelance}
\end{figure}

In our sample, all participants reported that AI was influencing their day-to-day work. Figure~\ref{fig:AIfreelance} shows the distribution of these ratings. Over half (57\%) reported that AI had a significant impact on their work in the past year (rating 5/5); the remaining 43\% rated the impact between 2 and 4 on a five-point scale (1 = no impact, 5 = significant). No respondent selected ‘no impact'.

To contextualize this impact,  we also posed an open-ended question inviting participants to describe their own perceptions of the positive or negative impact of generative AI on their freelance work. Through inductive thematic coding, we identified the main themes. Table~\ref{tab:ai-impacts} summarizes these findings, showing each type of impact, a brief description, the number of freelancers who mentioned it, and a sample quote from their responses.

The most frequently discussed positive impacts were \textit{Learning \& Skill Development} and \textit{Increased Efficiency \& Productivity}, highlighting AI’s role in accelerating skill acquisition and helping freelancers complete tasks more efficiently. Other positives included \textit{Improved Quality of Output}, where freelancers noted AI's ability to enhance their products; \textit{Creative Support \& Idea Generation} where they used AI as a brainstorming partner; and \textit{Expanded Opportunities}, where AI enabled them to offer new services. Across the corpus of responses, positive mentions of AI’s impact (124) slightly exceeded negative ones (110). Yet these benefits also came with some concerns: the most prominent negative impacts, such as \textit{Increased Market Competition, Over-reliance on Technology, and Skill Degradation}, indicate that freelancers see AI not only as a tool for greater productivity and skill growth, but also as a potential destabilizing force in their professional landscape. Additional concerns included \textit{Quality Concerns with AI Outputs}, \textit{Job Security Concerns}, and \textit{Cost Barriers}. 

\begin{table*}[htbp]
\centering
\caption{Positive and Negative Impacts of Generative AI on Freelance Knowledge Workers}
\label{tab:ai-impacts}
\resizebox{0.9\linewidth}{!}{%
\begin{tabular}{@{}clp{4.5cm}cp{5cm}@{}}
\toprule
\textbf{\#} & \textbf{Theme} & \textbf{Summary} & \textbf{Count} & \textbf{Representative Quote} \\
\midrule
\multicolumn{5}{l}{\textbf{POSITIVE IMPACTS OF AI}} \\
\midrule
1 & Increased Efficiency \& Productivity & Automation and speed for tasks like writing, coding, proposals, etc. & 29 & \textit{``AI has made work faster and more efficient. It helps me draft content, organize tasks''} \\
2 & Learning \& Skill Development & Helps learn new tools, coding, and improve professional skills & 30 & \textit{``Learning new skills using AI is easy and time efficient''} \\
3 & Improved Quality of Output & Polishes communication, improves deliverables, boosts client satisfaction & 25 & \textit{``It has saved me a lot of time. Has improved my communication skills''} \\
4 & Creative Support \& Idea Generation & Supports brainstorming, breaks creative blocks, offers varied ideas & 24 & \textit{``It allows me to get going with brainstorming rather than staring at a blank page''} \\
5 & Expanded Opportunities & Enables offering new services like AI-content or automation & 16 & \textit{``AI has enabled me to offer specialized services, like AI-powered solutions''} \\
\midrule
\multicolumn{5}{l}{\textbf{NEGATIVE IMPACTS OF AI}} \\
\midrule
6 & Increased Market Competition & Beginners using AI create more competition, potentially lowering rates & 33 & \textit{``Beginners can now use AI to perform tasks that once required deep expertise, sometimes lowering gig rates''} \\
7 & Unrealistic Client Expectations & Clients expect faster and cheaper work assuming AI replaces human expertise & 13 & \textit{``Sometimes it feels like clients expect everything to be done by AI, even the human touch''} \\
8 & Over-Reliance \& Skill Degradation & Heavy reliance on AI can lead to reduced confidence and creativity & 25 & \textit{``Sometimes I catch myself depending on AI for tasks I used to do confidently''} \\
9 & Quality Concerns with AI Outputs & AI output hallucinates and requires corrections & 10 & \textit{``I wasted time whenever the response had many hallucinations''} \\
10 & Job Security Concerns & Freelancers worry about being replaced or becoming obsolete & 18 & \textit{``I'm afraid of becoming obsolete, my client might prefer using AI instead of hiring me''} \\
11 & Cost Barriers & Advanced AI tools are expensive for some freelancers & 11 & \textit{``It is expensive to use better Gen AI''} \\
\bottomrule
\end{tabular}
}
\end{table*}

\begin{figure}[htbp]
  \centering
  \includegraphics[width=\columnwidth,
    alt={Horizontal bar chart of six AI use categories. Idea generation leads at 47 percent, roughly twice the next category; translation is shortest at 3 percent.}]
    {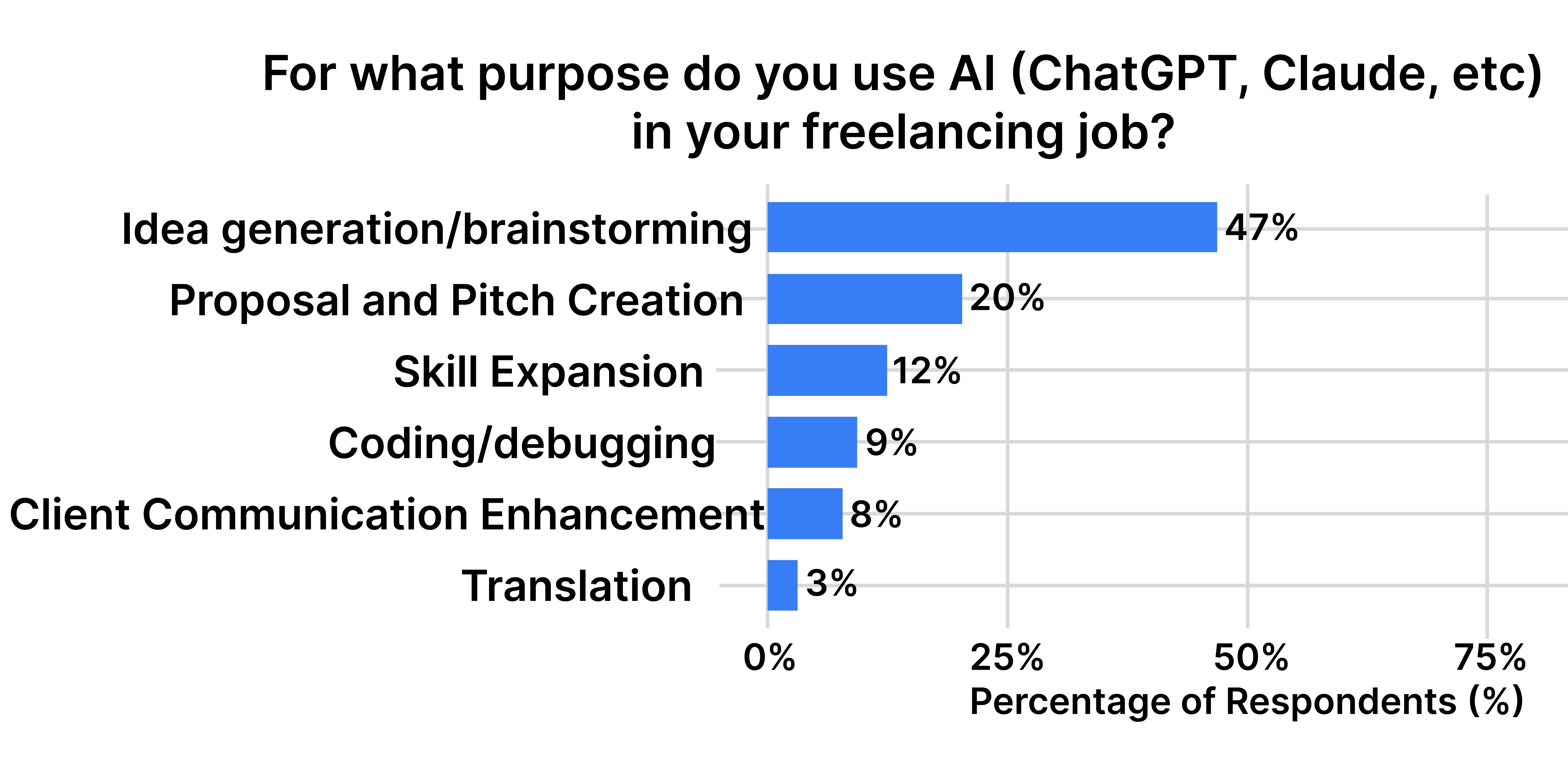}
  \caption{Self-reported purposes for using AI tools such as ChatGPT and Claude in freelance work.}
  \Description{Bar chart of AI use by purpose: idea generation and brainstorming 47 percent, proposal and pitch creation 20 percent, skill expansion 12 percent, coding and debugging 9 percent, client communication 8 percent, translation 3 percent.}
  \label{fig:useofAIchart}
\end{figure}

After identifying their perceptions of the impact of AI, we examined how freelancers actually used AI in their daily work. Note that Table~\ref{tab:ai-impacts} captures freelancers' perceived impacts of AI (what they believe AI does for them), while Figure~\ref{fig:useofAIchart} captures their reported uses (what they actually do with AI). The gap between perceiving AI as beneficial for learning (n=30, Table~\ref{tab:ai-impacts}) and actively using it for skill expansion (12\%, Figure~\ref{fig:useofAIchart}) became a focus of our interview investigation. Figure~\ref{fig:useofAIchart} illustrates the different use cases and the percentage of freelancers who selected each option from a select-all-that-apply item. The options were drawn from prior research \cite{Claudepaper}, though participants could also select ``other'' and provide additional examples. Nearly half (47\%) reported using AI for creative support, such as idea generation and brainstorming, followed by proposal and pitch creation (20\%). Individual freelancers also wrote in that they used AI to summarize content, manage projects, clarify job postings, communicate with clients, and cross-check their work.

The impact of a technology can also be measured by the time people spend using it. Freelancers in our study reported dedicating more than one-fourth of their work time to AI. When asked what percentage of their freelance work involved AI tools, most participants (about 70\%) said they used AI tools from one-quarter to three-quarters of their time. A smaller group (18\%) reported using AI less than one-quarter of the time, and 11\% mentioned using it more than three-quarters of the time.

While freelancers use AI to support their work, they also face barriers to its efficient use. When asked to select from a list of potential obstacles, the most frequently selected response was the rapid pace of change (20.5\%). This was followed by the cost of learning resources (19.2\%) and information overload (17.4\%). Other frequently mentioned obstacles included a lack of structured guidance (14. 7\%), time constraints (14. 3\%), and difficulty in identifying relevant skills (12. 1\%). Individual representatives also mentioned issues such as a lack of time, the release of new AI tools with varying capabilities, difficulty in identifying the right tools, and the presence of too many AI tools.

These diverse impacts create the context for our central focus, understanding how freelancers approach learning when faced with such disruption. We now turn to examining their upskilling patterns. 

\subsection{Patterns of Upskilling and Preferences for AI-Supported Learning}
 Freelance workers reported actively investing time in skill development. In our study 29.6\% reported participating in upskilling daily, and another 29.6\%  weekly. Additionally, nearly half (48\%) reported spending between 2-5 hours per week learning new skills for their freelancing careers, while 24\% reported spending more than 6 hours a week.  These patterns align with the Upwork Research Institute’s 2025 Future Workforce Index. The report notes that 32\% of skilled workers (freelancers and Full Time Employees) say they are “learning new skills all the time,” and that 87\% of freelancers (vs. 82\% of FTEs) learned a new skill in the past six months \cite{upwork2025}. 
 
 Choosing which area to focus on for upskilling is not random. When asked what skills they believed they needed to develop in the next 1-2 years, around 39\% of participants mentioned AI. 

In terms of which methods they choose for upskilling, in a single select-all-that-apply item, freelancers generally preferred self-directed approaches, choosing video tutorials (85.9\%; e.g., YouTube) and AI learning tools (70.4\%) over peer learning (64.7\%) and other community-learning methods. Nearly half of freelancers choose online courses (49.3\%), identifying specific programs such as Udemy, Coursera, skool.com, and LinkedIn Learning, which encompass both self-directed courses and community-learning programs. Finally, 42.3\% selected Industry Blogs/Articles, which is lower than the other categories but also shows value for learning from others, potentially more experienced freelancers. 
 
Professional certifications were also included in the response options for this item, to capture freelancers' interest in formal recognition, consistent with prior studies exploring how and why freelancers engage in professional development \cite{blaising2021making}. Although certification is not a direct method of upskilling, but rather a broader category encompassing credential-based learning practices, it was selected by 22 participants (31\%). This suggests that for roughly one-third of freelancers, obtaining certification is perceived as valuable. This emphasis on certification reflects an early indication of a broader challenge that recurs throughout our findings. While freelancers frequently described gaining practical skills through self-directed and AI-supported learning, these skills were not always legible to clients or platforms without formal credentials. As a result, participants often viewed certifications less as sites of learning and more as mechanisms for making AI-acquired capabilities visible and credible in the market. This tension indicates what we later conceptualize as invisible competencies: skills that freelancers genuinely develop, but struggle to signal or validate within competitive freelance labor markets.

\begin{figure}[!htbp]
  \centering
  \begin{minipage}[t]{0.48\textwidth}
    \centering
    \includegraphics[width=\linewidth]{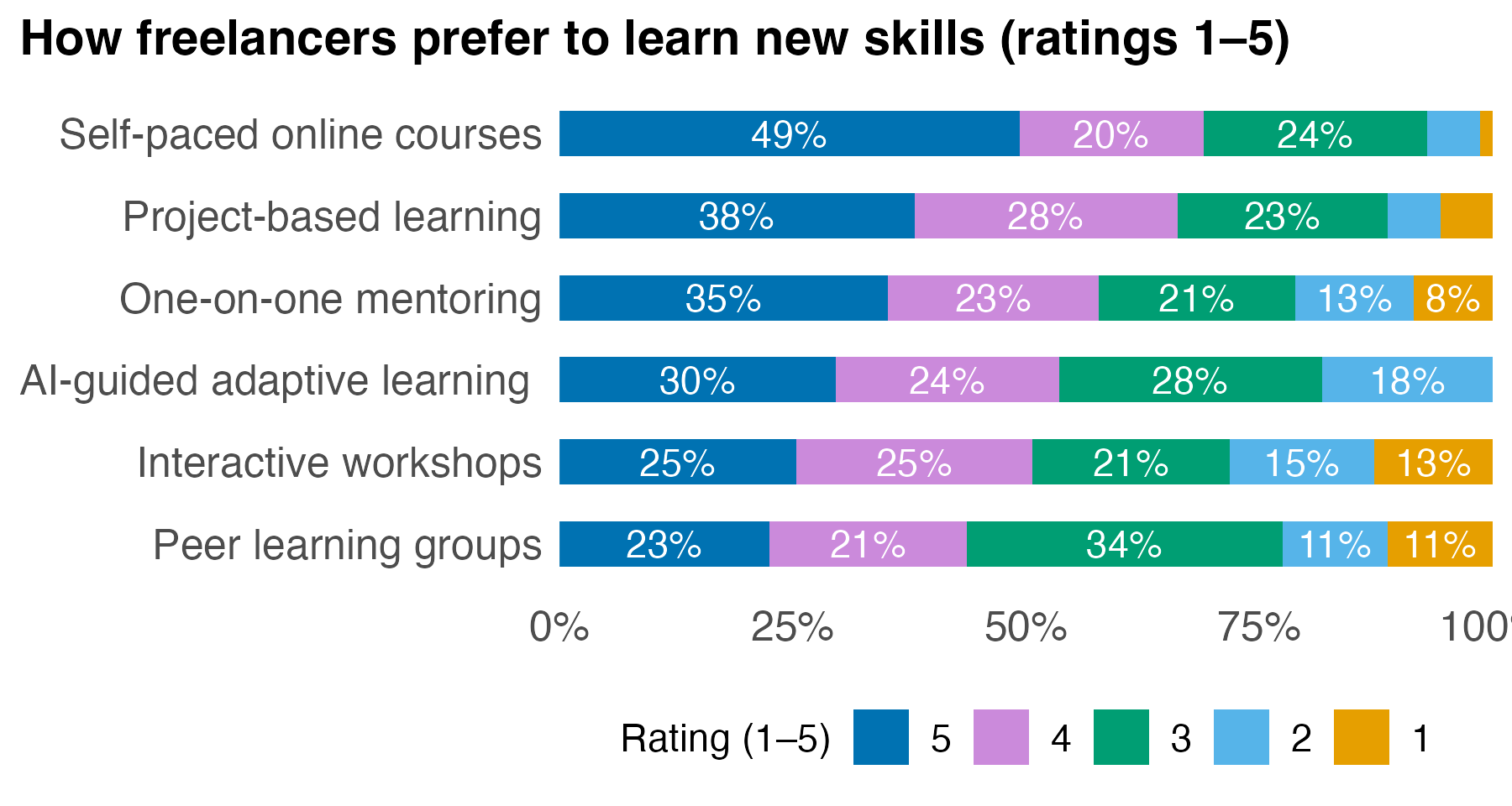}

    \Description{Stacked horizontal bar chart of learning method ratings. Self-paced online courses received the highest ratings; peer learning groups received the fewest ratings. AI-guided adaptive learning had no ratings of 1.}

    \caption{Preferred methods for learning new skills, rated from 1 (lowest) to 5 (highest).}
    \label{fig:preferred_upskilling}
  \end{minipage}
  \hfill
  \begin{minipage}[t]{0.48\textwidth}
     \centering
    \includegraphics[width=\linewidth]{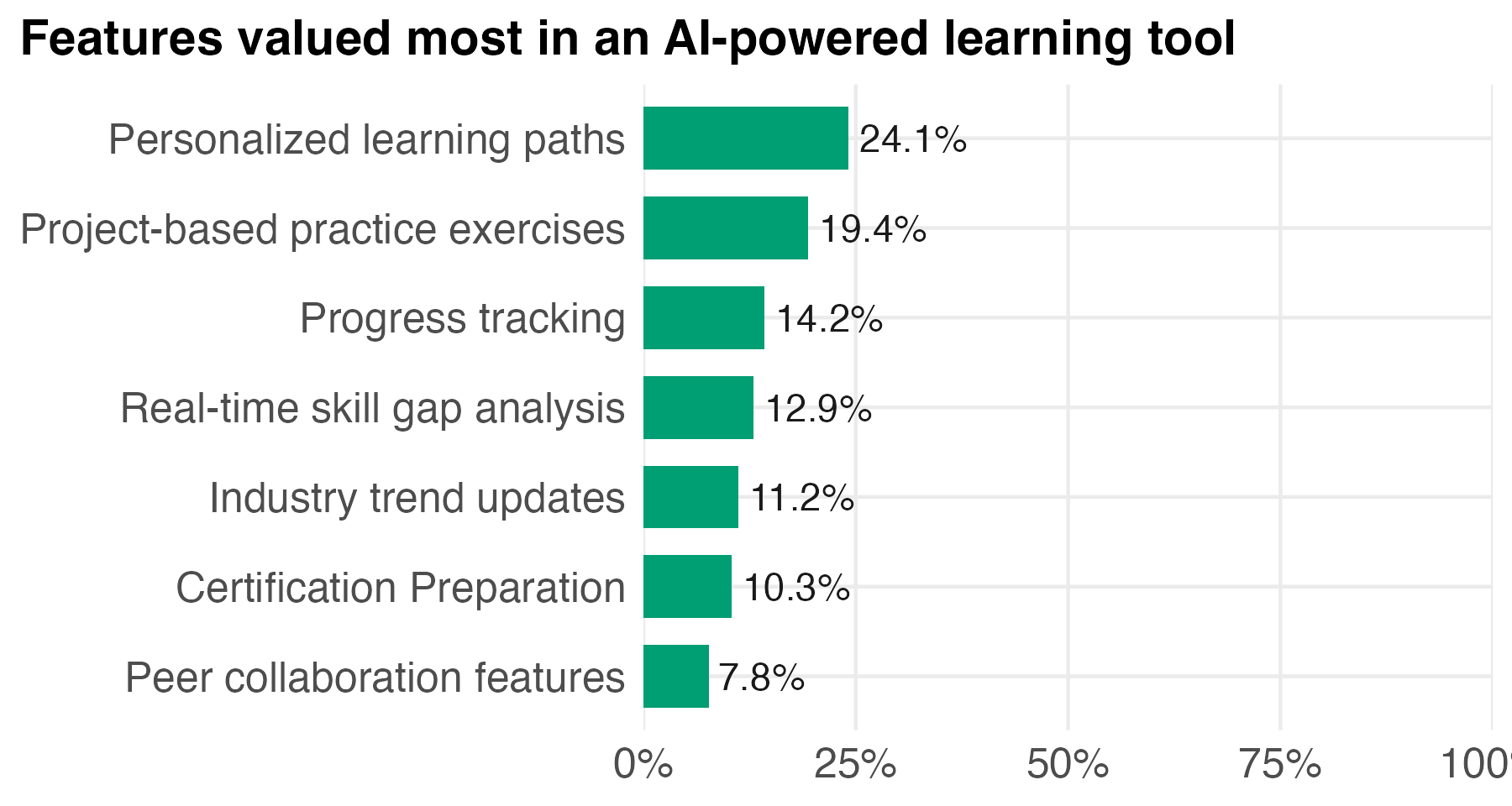}

    \Description{Bar chart showing top-three feature selections for an AI-powered learning tool. Personalized learning path is highest, followed by project-based practice exercises, and peer collaboration features is lowest.}

    \caption{Features valued most in an AI-powered learning tool. Respondents were asked to select their top three.}
    \label{fig:features_top3}
  \end{minipage}
\end{figure}

To compare actual practices with preferences, we asked freelancers to rate six learning methods on a 1–5 scale, where \textit{5} indicated a strong preference. We then calculated the percentage of participants rating each method a \textit{4} or \textit{5}. As shown in Figure \ref{fig:preferred_upskilling}, the most highly preferred approach was self-paced online courses, with about 70\% of respondents rating it positively, followed by project-based learning (66\%), underscoring a strong preference for flexible and hands-on formats. One-on-one mentoring (58\%) and AI-guided adaptive learning (54\%) were moderately favored. Although mentoring received more top scores than AI-guided learning, the latter showed broader acceptance overall, with higher mid-range ratings (\textit{2}s and \textit{3}s) and no respondents rating it as \textit{1}. Peer learning groups received the lowest endorsement (44\%). Overall, the distribution of responses highlights substantial variation in individual learning preferences across the sample.

Although AI-adaptive learning was not the most highly rated method used by participants for upskilling, when asked how interested they would be in using an AI-powered learning tool to master skills 82\% of participants responded very interested (13\% somewhat interested, 3\% not very interested, and 3\% neutral). A follow-up question was included to explore further what features participants would most value in an AI-powered learning tool. The most frequently selected features (Figure \ref{fig:features_top3}) included personalized learning path (24. 1\%), project-based practice exercises (19. 4\%), and progress tracking (14. 2\%). Followed by real-time skill gap analysis (12.9\%), industry trend updates (11.2 \%), certification preparation (10.3\%), and peer collaboration features (7.76\%).

In sum, freelancers are active upskillers who increasingly view AI as important for their learning needs. Yet, while 70.4\% of freelancers currently use AI tools for upskilling, this method ranks in the middle,  compared to project-based learning and self-paced online courses (see Fig. \ref{fig:preferred_upskilling}). However, more than 80\% of freelancers expressed a strong interest in an AI learning tool that would provide personalized learning paths and project-based exercises. This suggests that, despite their current use of AI tools for skill development, existing options may not fully meet their needs for personalization and practical experience.

\section{Interview Results}
We present the core themes that emerged from our qualitative analysis, organized into five subsections: 

\subsection{Drivers for Learning New Skills in an AI-Driven Labor Market}
Despite the growing influence of AI in reshaping digital labor across various fields \cite{Claudepaper}, we found that freelancers' upskilling efforts were not primarily driven by AI as a technological frontier. Instead, their learning remained anchored in client demands. However, AI amplified and destabilized this client-driven dynamic in ways that distinguish it from earlier cycles of tool adoption. For example, P16 explained that they had to learn various project management and collaboration tools because their clients expected them to use those platforms:  \textit{"The biggest challenge I face [in upskilling myself] is sometimes it's difficult to know what kind of tool is really relevant [to learn] because client A can say, 'We are using Monday.com [a project management tool],' and you learn Monday.com. Client B says, 'We are using Trello [another project management tool],' so you have to learn again. Then client C comes with Basecamp [another project management tool]. The client will not switch their subscription because of you. So you just have to keep learning the different tools."} 

Similarly, P5 explained that their learning was driven by market demand, specifically the types of jobs clients were requesting. Their motivation to upskill was not based on the release of new AI technologies. Instead, they would analyze the kinds of jobs clients wanted and then identify the skills they needed to complete those jobs more effectively. This client-centered approach guided their decisions about what to learn: \textit{"So this is a kind of a mix [my learning is driven by a kind of mix], a mix between market demand and whatever new skills I stumble upon during my work that I feel like if I learn them [learn the skills], it will make my work much easier. So, for example, I talked about web scraping. This was my first skill that I learned as a freelancer. And I learned it [learned the skill of web scrapping] because it was, it has, it had the highest demand [many clients were posting jobs related to web scrapping]."}

While client demand remained the primary driver of skill development, AI reshaped the conditions under which freelancers interpreted and responded to those demands. Freelancers described AI as a background condition that intensified uncertainty about what to learn and how long any given skill would remain valuable. The pace of AI tool proliferation created a form of decision fatigue distinct from earlier periods of technological change: freelancers were contending with a rapid and unpredictable expansion of the tool landscape itself, not just learning new tools as clients adopted them. This stress came less from the tools themselves than from the fear that clients would soon expect mastery of rapidly evolving AI technologies. Consequently, freelancers often felt overwhelmed by the constant pressure to stay current and worried that skills they had invested time in learning might quickly become obsolete. As P4 emphasized: \textit{``I have felt very overwhelmed because...I may be trying to master one skill on Google Ads. But, even before you can master everything [on the AI-driven platform of Google Ads], TikTok Ads is already there...Then there are X Ads, LinkedIn Ads. So it can be so overwhelming because you're trying to ask yourself, where should I focus? And to have a competitive edge, you need to have one focus, one niche. And sometimes the industry evolves so fast that some of these skills end up becoming obsolete..."}

This anxiety went beyond general concern about technological change. Participants recognized that AI could create new skill requirements and devalue existing expertise at the same time.

\subsection{Leveraging AI to Structure Learning Goals and Integrate Them into Workflows}
Our survey revealed that freelancers were beginning to use AI tools to support their learning and skill development, the interviews offered deeper insight into how freelancers actually did this. A common approach was using AI to create personalized learning roadmaps. Freelancers described asking AI to break down complex learning goals into smaller, manageable steps tailored to their specific needs. They also used AI to help turn these small learning tasks into actions they could incorporate into their daily work. Instead of treating upskilling as a separate effort, freelancers used AI to integrate learning into their regular workflows in a practical and sustainable way. Participant P7 described this process: \textit{"So, the first process 
is that I use AI to brainstorm what to learn...after using AI to brainstorm, I tell the AI system I use, to give me a detailed and organized checklist, breaking everything into a step by step section of how I will go about learning each topic leading to a skill. That means the AI is going to break everything down for me into segments that I can take my time every day to learn."}

Part of the reason freelancers used AI to structure their learning process and find ways to integrate learning into their work schedules was that they typically juggle multiple activities as part of their freelancing responsibilities. AI thus became essential in helping them organize their learning and seamlessly incorporate it within their busy work schedules. P1 shared some of this reasoning: \textit{"Sometimes it gets overwhelming when you have to juggle between your timing, workload and learning. I leverage some AI tools to help me to learn faster. I do, I use AI to structure my learning. I use them to prioritize and structure my day for a particular period of time"}.

Given that freelancers valued AI tools that could help structure learning into their daily workflow, participants also imagined future AI systems that could track their progress in learning a skill and then match them with jobs or projects to help them practice or further develop that skill. For example, participant P5 described the following: \textit{"For the more abstract topic, practical experience is in my opinion, the only way to learn. Because an explanation video, as I said before, makes you better understand what this topic is, but not actually how to do so. So if an AI agent for example, generates a problem for me, generates a project for me to do or finds a gig about it, and then guide me through this project, if I am stuck on certain parts, then this will certainly help"}.

This strategic approach of using AI to structure learning within their everyday work schedules extended to using AI for targeted skill acquisition and quick concept clarification. P2 described this approach: \textit{"...I am actually using, using ChatGPT to just be like, hey, I'm learning Go [a programming language]. Give me a rundown of the syntax [structure of the programming language]. And then it does that [gives them an overview of the programming language and its syntax]"}.

Freelancers also described using AI to help them structure their learning in ways that better matched their preferred learning styles, even when these styles could be relatively complex. One participant (P6) articulated their approach: \textit{"...and something that I do like with ChatGPT and learning Spanish [a language they were learning  to get new gigs], [is that] I am very aware of how I learn best based off of experimenting a lot... so, I know systems and structures that help me focus well. And so with this, I've asked it [AI] to develop different, like, to develop the optimal learning strategy for me for learning Spanish"}.

Overall, we found that freelancers primarily used AI to structure their learning process, making it easier to integrate into daily workflows. This AI-driven structuring also enabled them to tailor learning to their individual styles and focus on specific skills they wanted to develop.

\subsection{When AI-Acquired Learning Does Not Translate to Recognition}
Our survey revealed that while freelancers increasingly incorporated AI tools into their learning practices, these tools were not their primary learning resources. Through interviews, freelancers explained why AI remained secondary as it introduced additional effort to interpret, evaluate, and verify learning content, increasing freelancers’ learning burden. Participants described needing to assess credibility, reconcile conflicting guidance, and translate AI-generated content into practice, often under time pressure.

Freelancers frequently reported feeling overwhelmed by the volume of AI-generated learning content. Although generative AI tools offered extensive learning resources, this abundance often produced decision fatigue, making it difficult for participants to evaluate and select appropriate materials. 
P7 described this fundamental challenge: \textit{"I think it all comes down to, you know, it comes down to resources...overwhelming number of resources when I'm trying to learn. There are a lot of information out today and most of them are not verified"}. Freelancers in our study highlighted another layer of this overload, particularly the proliferation of AI learning tools. P3 mentioned: \textit{"Some of the concerns I have with AI-powered learning tools are that there are so many of them to learn from. Sometimes I don't know which one. There are so many of them. You wake up each morning, and there's one improvement over the previous one. Sometimes it could be overwhelming. You don't really know which one to focus on. If you're not careful, it could be distracting."} As a result, learning using generative AI becomes less about engaging with content and more about navigating an unstable ecosystem of resources.

In addition to information overload, AI-generated learning content introduced a new challenge: inconsistency and unreliability. Participants noted that AI tools often gave varying answers to similar questions, making it hard to know which learning path to follow. As P4 noted, \textit{"..It's [the AI is] never consistent. So let's say today I've asked it [the AI to] give me a roadmap on my digital marketing apps skill journey. It will give me one [roadmap]. But tomorrow I ask it the same question, it will give me a different response, and if I ask it the next day, a different response. So you ask yourself, okay! Which response should I focus on?"}. This inconsistency extended beyond just varying responses as AI-generated content often failed to work in practice, forcing freelancers to debug and troubleshoot rather than learn. P5 described this frustrating experience: \textit{"It happened to me a lot before when I was learning a new skill. For example, I have a question that I can't find an answer for. I will ask this gpt, for example, or Google Gemini for an answer and he will give me an [unreliable] piece of information, for example. I try to run it and I find half the libraries are deprecated."}. 

Inconsistency and unreliability in AI-generated learning content became a burden for freelancers, as it forced them to validate and debug the information instead of focusing on learning. They had to develop evaluation skills to distinguish useful content from unreliable advice, adding to their workload. As P6 noted, freelancers not only had to learn new skills but also assess the quality of AI-generated information: \textit{"I think that there is a sense of responsibility of self fact checking [skills related to evaluating whether the AI suggestions are useful] right now. Because, for example, when I am trying to gather learning resources, it [the AI tool] provides great resources like the titles of resources, but I need to go and fact check that these resources actually exist and are credible"}.

Beyond these challenges, freelancers emphasized that AI-supported learning rarely translated into legible signals of competence for clients. While participants described acquiring valuable skills through AI-mediated learning, they lacked credible mechanisms to demonstrate or validate these capabilities within platform labor markets. Clients often valued recognized credentials and prior experience as indicators of expertise.

P4 emphasized the importance of certification from the client perspective: \textit{'So one of the things they [Clients on freelance platform] do value is experience. If I'm able to show a client my portfolio and work I've done before, this is a bonus. And they also feel more confident whenever I show them my certifications because once they see that, ah, Google has certified you, clearly you can do my work."}

In contrast, AI-based learning frequently produced skills without comparable forms of external recognition. This resulted in what we term "invisible competencies", skills that freelancers genuinely acquire through AI-mediated learning but cannot easily signal or validate to clients. Rather than reflecting a simple absence of certification, participants described a structural mismatch between learning and legibility, where learning pathways evolve faster than the mechanisms platforms and clients use to evaluate competence. As P3 articulated: \textit{"Sometimes you learn to learn and no certificate to back it up. Some clients will want to see if you claim you have learned this and so stuffs, where is a certificate to back it up? But, sometimes the platforms [AI learning platforms] don't have a certificate to back it up [to show a certification that you gained the skill]..."}.

This challenge was distinct from the difficulty of signaling informally acquired knowledge more broadly. When freelancers learned from peers, they could reference who they learned from and draw on that person's professional credibility as indirect validation. AI-acquired skills, by contrast, arrived without any attributable source. Participants described having no equivalent way to explain to clients how they developed a capability when the learning occurred through private conversations with Generative AI.

Several freelancers described enrolling in formal courses primarily to obtain recognizable credentials, even when those courses did not substantially differ from what they could learn through AI tools. P19 explained this strategy: \textit{'One of the things I sort of do is try to get certified.... I take a lot of classes that are something that I'm really interested in and then I just go through and take those classes as well."} This behavior highlights how freelancers strategically pursue credentialed learning pathways not solely for skill acquisition, but to obtain market-recognized signals that make their upskilling legible to clients.

\subsection{Learning with AI and Peers}
Our survey uncovered that freelancers practiced peer learning and AI tools to a similar degree. In our interviews, freelancers consistently described turning to peer learning to address limitations of current AI tools, particularly when learning required contextual judgment, experiential grounding, or clarification amid uncertainty. P7 emphasized the importance of peers in navigating confusing or overwhelming AI-generated learning resources: \textit{"Most of the time we are confused by our AI learning resources. We need someone in that position [someone who is also a freelancer] to come in and teach us. So having a community of learners and experts within our same place [learners and experts who are also freelancers], they guide you by the hand, taking you through whatever it is you're finding difficult, helping you to up-skill. Basically, we [freelancers] help one another wherever we find difficulties."} Across interviews, freelancers described using AI tools for structured, factual, or low-stakes questions, while relying on peers for guidance rooted in lived experience and domain-specific context that AI could not provide. 

Critically, freelancers emphasized that the value of peer learning stemmed not only from personalization but from learning from identifiable individuals whose experience, context, and professional outcomes were known. This allowed freelancers to assess the relevance and trustworthiness of shared knowledge without the extensive verification that AI-generated content often required, directly addressing the verification burden described in the previous section. Because AI learning content arrived without provenance (no indication of source, validity or basis for recommendations), peer learning provided traceability: who shared the knowledge, under what conditions, and with what results. P6 described this approach: \textit{"If I only have basic level questions and I feel like that there's enough information online... then  I am not going to ask these professionals or people that have been there [those basic level questions]. Instead, I can ask them something that's actually deeper, something that has some complexity [...] I'll ask them  a more complex question and that's something you can ask a human about based off of their experience."} Such an approach addressed the verification burden described in the previous section. 

The value of learning from identifiable, experienced peers became particularly evident when freelancers described gaining professional benefits through peer learning, such as accessing specialized knowledge and securing new opportunities. P20 recounted how collaborating with peers directly led to business success: \textit{"One of my fellow freelancers that I am close with, he is in HR [he is a freelancer who work in human resources], but he is in like the HR business partner space... he was able to kind of walk me through some like state regulations pertaining to human resources, which was really nice of him. And I ended up getting that gig [a gig related to HR and state regulations] and was able to help that client"}. Here, the knowledge was valuable precisely because P20 knew its source: a practicing HR professional whose expertise could be verified through their ongoing work. Similar regulatory information from a generative AI tool would have required independent fact-checking due to the absence of source attribution.

Peer learning created supportive environments that helped freelancers learn and progress in their gigs. P14 highlighted how this dynamic supported both learning and work advancement: \textit{"We actually grouped for a project [gig]. We are like 40 members. And the reason why we went as far as we did was because we were all there. We are like holding each other up as peers. We are actually called peers in the community.... it actually helped us in the days where one peer might not be feeling up to it [to finish work related to the gig]. Others would just encourage this person to, like, keep going. Or at times, some people [freelancers] might not really understand what is going on. The other peers will just take over, like, make you see a better point of view or just teach you for you to understand. So peer learning is actually great..."}.

However, access to peer learning was uneven, creating barriers for freelancers without strong networks and potentially pushing them toward AI tools that, while accessible, lacked attribution benefits. P8 noted that lack of access to experienced mentors particularly disadvantaged newer freelancers: \textit{"I feel like if you have a mentor in a particular field, you know, the mentor has gone ahead of you. The mentor has seen some of these challenges and problems and difficulties.. the mentor helps you navigate your way properly. So one of the challenges that I feel like virtual assistants [a type of freelancer] face is accessibility to mentorship, which will in turn impact on how much resources they can get access to"}.

We observed that freelancers not only decided when to engage in peer learning but also managed what to share and under what conditions. Many took a strategic approach, balancing helpfulness with protecting their competitive edge. P5 articulated this nuance:  
\textit{"It is usually when both people stand to benefit from the exchange [that we engage in peer learning]. I am usually willing to share my knowledge, but at the same time I am a very competitive person. So, if I know that this person [freelancer] in front of me is an actual competitor and not just someone asking me about a question, then usually, I will not be inclined to share [share knowledge], to give away my advantages for free. But if they are willing to share some of their specialized knowledge in exchange, in exchange for my own [my own specialized knowledge], then this becomes a kind of trade and I am willing to exchange to participate in these exchanges."}

Freelancers’ selectiveness in sharing knowledge often stemmed from the competitive nature of freelancing, which posed challenges to peer learning.
As P20 observed: 
\textit{"The competitive nature of the gig economy may deter freelancers from befriending others in the same field due to direct competition for opportunities."}  

Freelancers often drew clear boundaries between different types of information. For example, many were willing to provide general guidance but not share proprietary work products. As P5 further explained:  
\textit{"So if someone just wants the final product from my previous projects, for example, I will usually tell them no. But I am willing to give them a general outline of the process I went through and point them in the right direction."}

The level of openness in peer learning also depended on the context and audience. Freelancers were more inclined to share in trusted, private settings than in public forums. P18 described this distinction:  
\textit{"I prefer to share in private [private groups], because these are the people that I know more of. I know... maybe I can say I know every one of them [know the freelancers in the private group], or I know most of them. I'm open to that, private groups [open to doing peer learning in private groups]. But in the public [doing peer learning in public online spaces], where your information can be accessed by anyone, I'll keep it brief and maybe more professional so that I don't spill more than I should."}

Taken together, these findings reveal a fundamental tension in freelancers’ learning ecosystems. Peer learning offers contextual grounding and credibility that AI tools often lack, helping freelancers manage information overload and reduce verification effort. Yet competitive pressures and uneven access constrain when and how freelancers engage in peer-based knowledge exchange. As a result, freelancers strategically navigate between AI tools for private, exploratory, low-stakes learning and peer learning for contextual validation, opportunity access, and legitimacy-building.

\subsection{Envisioning Generative AI-powered Self-Directed Learning System}
Through our interviews, freelancers shared their visions for the future of AI systems that could support their learning. They emphasized the need for AI systems that can actively manage cognitive load by filtering learning-related information, adjusting content difficulty, and providing targeted feedback. For example, P3 described such a system: \textit{"This adaptive learning [AI learning system] provides me timely feedback, helps me adjust to difficulty levels, and it offers me some personalized resources to learn from, like videos, articles, and all that"}.  

This emphasis on personalized resource curation reflects freelancers' desire to take control of their learning environment, a core component of self-directed learning \cite{garrison1997self}. Participants specifically wanted AI systems that could reduce extraneous cognitive load by eliminating irrelevant information and focusing attention on essential learning elements. As P10  articulated, \textit{"if it [their proposed AI learning tool] gives me personalized options, like, if I want to learn a specific thing, and the AI tool gives me the exact specific things and no other, like, no filler, but only the knowledge that I need for the specific topic or skill, I would really want such a tool...".} 

Similarly, another participant (P14) emphasized the importance of giving learners control over personalization, enabling them to better manage their cognitive load: \textit{"I would like to cut it off or shift it [cut off or shift the information the AI learning tool provides to them] and just move to what I need at that moment [move or control the information the AI learning shows to whatever they need at the moment]. So, I feel it would be really great if it [AI learning tool] were personalized. The personalized view feature can be introduced into any tool for gig workers to manage their mental energy!"}. This vision reflects the understanding that effective scaffolding must be responsive to learners' immediate cognitive capacity and goals.

Freelancers envisioned AI learning tools that could support self-monitoring through demonstration, providing examples, and guided practice, allowing learners to observe professional procedures and evaluate their own progress. This type of support could help address the gap between theoretical knowledge and practical application that freelancers frequently encounter. Participants wanted AI tools that could demonstrate professional practices through concrete examples. P16 emphasized the importance of demonstration: \textit{"It [AI learning tool] should try to give examples because this is what helps. For example, if you are talking about how to write a proposal or how to stand out, try to [the AI tool should try to], like, show..., give relevant examples of profiles that are standing out. That way, when I take a look, I kind of know that, okay, this is how to set my own up."}
Participants also envisioned a Generative AI tool that could learn each learner's unique patterns and tailor guidance accordingly. This would support the self-monitoring process by helping freelancers track how they learn best. P9 explained: \textit{"I would be willing to use an AI-powered adaptive learning tool if it will definitely learn my learning pattern, the way I deal or the way I, you know, search through things... and it would help me by copying the pattern or copying the procedure for me and giving me advice based on my learning pattern..."} Others envisioned AI that could temporarily complete steps for the learner, then gradually reduce support as competence develops. As P10 said: \textit{"It [AI learning tool] should have the ability to make decisions for them [for the learner] and complete tasks for them [for the learners] instead of just making suggestions. This would be based on, like, [learners'] previous experience and personalized adaptiveness..."}
These visions reflect freelancers' awareness of their learning challenges: information overload from current AI tools, the gap between theoretical knowledge and practical application, and the need for personalized guidance that supports autonomous skill development.

\section{Discussion}
Our mixed-methods study reveals a set of tensions in how freelancers upskill using current generative AI tools. Freelancers increasingly adopt these tools and express a strong interest in leveraging generative AI for learning, but they often introduce additional burdens, including information overload and the need to verify outputs. Drawing on Garrison's self-directed learning theory \cite{garrison1997self}, we argue that market pressures shift freelancer learning away from growth-oriented goals toward what we call \emph{learning-as-survival}, where upskilling is driven by the need to remain competitive and secure income. We also identify \emph{invisible competencies}: freelancers acquire skills through generative AI use but struggle to demonstrate or validate them in freelance labor markets.

\subsection{Market-Driven Learning: From Growth to Survival}
Our survey found that 39\% of participants consider learning AI-related skills a top priority for upskilling over the next 1–2 years. However, our qualitative data suggests that freelancers’ engagement with AI for learning is primarily driven by client demands. For many freelancers, upskilling in AI seems less about embracing a new technological frontier and more about meeting explicit expectations for AI-augmented deliverables in a highly competitive market.

This market-driven orientation reflects a reactive mode of professional development. Freelancers often described adapting their skill set not through strategic long-term planning but as a necessary response to clients' evolving demands. This highlights the complex nature of productive client-freelancer relations, which has an impact on freelancers' professional development \cite{blaising2021making, venkatesh2023measure}. As such, freelancers framed their AI upskilling efforts as strategies for economic survival, rather than as proactive career investments or expressions of personal interest in the AI-enhanced technologies. This pattern marks a shift from traditional adult learning models centered on self-directed growth \cite{garrison1997self}, towards what we term  \emph{``economically constrained self-direction''}, where market imperatives appear to shape both the choice and perceived achievability of learning goals.

Using Garrison’s self-directed learning framework \cite{garrison1997self}, our findings show that freelancers’ motivation to learn is shaped by context, particularly economic precarity and client-driven expectations. In this framework, self-directed learning involves taking responsibility for one’s learning through three connected processes: self-management (planning and organizing learning activities), self-monitoring (evaluating progress and adjusting strategies), and motivation (sustaining effort and maintaining goal direction). Applied to our results, the model suggests that motivation is not only an individual disposition. It is also a response to external pressures. Because freelancers contend with unstable income and must meet client expectations, their learning priorities often emphasize immediate market viability over longer-term professional growth.

In line with this, freelancers in our study demonstrated self-management, often structuring learning around active client projects, and self-monitoring, for example, by tracking market trends and perceived skill gaps \cite{wei2024person, hsieh2025gig2gether}. However, their learning goals were frequently constrained by how they interpreted shifting market signals and client demands. These interpretations were shaped by personal experience and local context, which could be incomplete, short-lived, or only partially accurate. They also differed across freelancers depending on access to stable work, professional networks, and prior exposure to market information. This raises equity concerns about who can make well-informed learning decisions in a rapidly changing environment. At a collective level, if the majority of freelancers remain locked in survival-mode learning, the platform labor market risks a narrowing of skill development toward short-term client needs, with freelancers who lack financial stability or strong professional networks disproportionately unable to invest in the longer-term, exploratory learning that could broaden their career trajectories.

These pressures were further intensified by platform dynamics. Algorithmic reputation systems create reputational overhead that influences how freelancers approach skill development \cite{zhang2022algorithmic}. Many freelancers described avoiding experimentation with unfamiliar skills in client-facing work because visible mistakes could harm ratings and future opportunities. This produces a persistent tension between the learning value of experimentation and the reputational risks of platform-mediated labor \cite{wood2019good}, reinforcing reactive, client-driven upskilling over exploratory learning. 

This dynamic is further complicated by a tension specific to AI-mediated learning: generative AI is both the source of the disruption freelancers are responding to and the tool they use to respond. Freelancers described feeling overwhelmed less by the tools than by the expectation that clients would soon demand mastery of rapidly evolving AI technologies, producing anxiety about whether current learning investments would retain value. As a result, freelancers found themselves dependent on the very technology that threatens their market position, reinforcing the survival orientation rather than enabling growth-oriented exploration.

\subsection{Generative AI Shifts Learning Effort and Produces Invisible Competencies}
Our findings reveal how freelancers navigate tensions between learning, verification, and market legibility when using AI learning tools. While 70.4\% of participants actively used AI tools for upskilling and 82\% expressed strong interest in AI-powered learning systems, AI learning tools were not among their most preferred methods for learning new skills. Although freelancers saw clear opportunities in these tools, they also described persistent challenges, particularly inconsistency and unreliability. These issues create what prior research characterizes as a verification burden \cite{VerificationofLLMLabels, ChallangeinLLMforlearning}. Unlike traditional educational resources that typically undergo editorial review \cite{conole2005learning, stein2001textbook}, AI-generated content often requires continuous fact-checking by individual learners, which can effectively increase their learning workload \cite{toxtli2021quantifying}.

Freelancers also reported that AI-based learning can produce a “credentialing gap,” in which workers develop valuable skills but lack formal recognition of those skills \cite{lang2023workforce}. About one-third of our participants identified this as a key limitation of AI-powered learning tools, emphasizing the need to supplement AI learning with formal courses to obtain recognized certifications. This gap can be particularly consequential in platform-mediated freelance markets, where credentials operate as trust signals that help clients decide which workers to hire for specific tasks \cite{spence1978job,pallais2014inefficient}. Without verifiable proof of AI-acquired skills, freelancers face what we term \emph{``invisible competencies'',} meaning real capabilities that are difficult to communicate to potential clients.

This challenge differs from other forms of informal learning. Skills acquired through peer exchange carry a social trace: the learner can reference who taught them, and that person's reputation serves as indirect validation. AI-mediated learning leaves no equivalent trace. It happens privately, produces no relationship with a mentor or peer that others can verify, and is associated with a technology that clients may distrust. The resulting skills are not just uncredentialed but lack the social context through which informal learning is usually recognized.

This finding extends prior work on verification burden by showing that AI-mediated learning produces not only increased verification effort, but also a distinct form of market invisibility. This invisibility can undermine the motivation component of Self-Directed Learning, as freelancers may question whether time invested in AI-assisted learning will translate into market value. In this sense, the credentialing gap is not simply a documentation issue. It can be a structural barrier that shapes which learning paths freelancers are willing and able to pursue.

\subsection{Peer Learning as a Source of Traceability in AI-Mediated Upskilling}
A key insight from our study is that peer learning provides a form of traceability that current AI learning tools often lack. Knowledge shared by peers is grounded in identifiable individuals with known experience, context, and professional trajectories. This allows freelancers to assess credibility without extensive verification. This finding aligns with theories of situated knowledge \cite{haraway2013situated}, which emphasize that the trustworthiness of knowledge is inseparable from the social and professional conditions under which it is produced. In this way, peer learning can reduce the verification burden associated with AI-generated learning content.

At the same time, our findings reveal that freelancers operate within a complex ecosystem of interconnected barriers that constrain effective peer learning. While prior work has documented structural barriers such as time scarcity and competitive pressures in freelance work \cite{gray2019ghost}, our study shows how these barriers become particularly salient in learning contexts. Freelancers described competitive environments in which knowledge sharing carries strategic risks, which contrasts with peer-learning models that assume open and reciprocal exchange for collective benefit \cite{qiu2025self, blaising2021making}. Although 64\% of participants reported already engaging in peer learning for upskilling, they rated it poorly as a preferred learning method. This tension reflects freelancers’ awareness of the risks involved, which our participants expressed include: oversharing, loss of competitive advantage, and the need to carefully manage professional boundaries.

Despite these concerns, freelancers also emphasized that peer learning remains essential for developing specialized knowledge, accessing opportunities, and establishing professional legitimacy. This tension points to a design challenge: enabling knowledge exchange while protecting individual competitive advantage. In this context, freelancers increasingly described generative AI as a lower-risk complement to peer learning. AI tools allowed them to explore new skills privately and iteratively, while reserving peer interactions for validation, opportunity access, and signaling credibility.

Rather than viewing generative AI as a replacement for peer learning, participants envisioned AI as a potential mediator of peer-mediated learning that could help address the isolation of freelance work \cite{pea2018social, imteyaz2024gigsense, seetharaman2021delivery}. Freelancers imagined AI systems that could support matching with peers who have complementary, rather than directly competing, expertise, enabling mutual learning benefits while minimizing competitive risk.

These findings align with Self-Directed Learning theory, which emphasizes that learning is not purely individual, but embedded within social and contextual conditions \cite{morris2019, neureiter2017exploring}. Garrison’s framework highlights that while learners take responsibility for directing their learning, social interaction plays a critical role in confirming, validating, and contextualizing newly acquired knowledge \cite{garrison1997self}. This was evident in our qualitative data, where freelancers consistently pointed to peers as crucial sources of opportunity, specialization, and legitimacy. Designing safer peer-learning environments that balance knowledge exchange with protection of competitive advantage can therefore strengthen trust and participation in peer learning, while complementing the role of generative AI in freelance upskilling.

\subsection{Design Implications}
Building on our findings, we propose a set of design recommendations for generative AI–based upskilling tools that address learning under market pressure, verification burden, invisible competencies, and competitive peer learning dynamics. 

\textbf{Market-Informed Upskilling Tools.} Our findings show that freelancers are effective at tracking their learning activities, but they often struggle to translate reactive, client-driven needs into long-term strategic growth. We propose tools that integrate real-time signals of client and market demand directly into freelancers’ existing workflows and automatically cluster client-requested skills into coherent skill domains over time. For example, a freelancer who learns “Excel data analysis” for one client, “Tableau visualization” for another, and “Python basics” for a third could see these skills grouped as progress toward a “Data Analytics Consultant” position. By surfacing these patterns, such tools would help freelancers turn scattered, reactive learning into coherent upskilling trajectories, strengthening the self-management component of Self-Directed Learning \cite{garrison1997self}.

\textbf{Verification and Skill-to-Portfolio Translation.} To address both the verification burden imposed by AI-generated learning content and the problem of invisible competencies, we propose integrated systems that support both the validation of learning materials and the translation of AI-acquired skills into market-legible evidence. Such systems could cross-reference AI-generated learning paths with verified sources, display confidence scores, and flag inconsistencies across industry blogs, course platforms, and job postings. Building on this verified learning trace, the system could then document skills acquired through AI guidance using timestamped evidence, generate portfolio artifacts that demonstrate competency, and create “skill stories” that explain how skills were developed and applied. By linking verification directly to portfolio construction, these systems would reduce learning burden while supporting the motivation component of Self-Directed Learning \cite{garrison1997self}, helping freelancers trust that their learning investments can translate into recognized market value.

Such systems could extend the traceability that freelancers associate with peer learning to AI-generated content. One approach is linking learning path recommendations to  outcome data from freelancers who followed similar trajectories, such as the proportion of freelancers in a given domain who successfully applied a skill after a particular sequence. Another is community annotation, where freelancers who completed a learning path can flag what worked and what did not in practice. This would give AI-generated guidance the kind of experiential grounding that freelancers currently find only through peer interactions.

\textbf{AI-Mediated Peer Learning with Competitive Protection.} To address the tension between the benefits of peer learning and the competitive risks it can create, we propose two complementary features. First, Complementary Skills Matching would connect freelancers with peers whose expertise is complementary rather than directly competing. By analyzing skill profiles, market positioning, and learning goals, the system could form partnerships where both parties benefit while minimizing the risk of losing client opportunities. Second, we propose AI-to-Peer Validation, a hybrid support model that integrates AI-assisted learning with targeted human feedback. In this model, a freelancer may use generative AI tools to advance along a learning path, but when the system detects uncertainty, contradictions, or context-sensitive decisions, it could route the freelancer to relevant human peers who can validate, interpret, or critique what the AI suggests. This is especially important when assessing whether an AI recommendation is appropriate for the freelancer’s specific work context, clients, and platform norms. Such a system could be implemented through paid or volunteer-based participation \cite{FutureofCrowd,flores2020understanding}, where freelancers build social capital by supporting one another \cite{gray2016crowd,rzeszotarski2014estimating}. This design direction could align with prior approaches in friendsourcing and learnersourcing \cite{bernstein2008personalization,bernstein2010crowd,glassman2015learner,kim2015learnersourcing}, leveraging structured peer contributions to provide timely, contextualized assistance. Beyond validation, such systems could support safer experimentation by creating simulated practice environments that mirror real platform conditions: mock client briefs, realistic project constraints, and peer review, without actual reputational consequences. This would allow freelancers to apply newly learned skills, receive feedback, and build confidence before deploying them in client-facing work.

\subsubsection{Limitation and Future Work}
Our study has several limitations. Participants were recruited exclusively from Upwork, which, despite its size and global reach, has particular algorithmic management practices, incentive structures, and reputational systems that may not generalize to other platforms. Our sample also skews toward knowledge workers who already use AI tools for learning; those who are reluctant to adopt generative AI may have different adaptation strategies worth examining. At the same time, focusing on early adopters remains valuable, as it provides some of the first empirical insights into how knowledge workers on freelance platforms are upskilling in the emerging landscape of generative AI–mediated work. The survey analysis is descriptive, reflecting both the sample size (n = 71) and its exploratory purpose. Larger samples would allow for inferential analysis of relationships between freelancing experience, AI usage, and learning preferences. Our findings also rely on self-reported practices, which may not reflect actual skill acquisition or performance. Controlled comparisons measuring concrete outcomes, such as client acquisition or earnings when using generative AI for upskilling, would strengthen these claims. Finally, we focus on learning and upskilling without attempting to characterize broader labor market shifts, power dynamics, or long-term career trajectories shaped by generative AI.

\section{Conclusion}
This paper examined how freelance knowledge workers use generative AI tools to support skill development in competitive, platform-mediated labor markets. Freelancers increasingly turn to generative AI to structure learning and adapt to shifting client demands, but these tools introduce verification overhead, limited contextual relevance, and a credentialing gap that produces what we call invisible competencies. AI-acquired skills lack the social traces that make peer-learned skills legible to clients, which shapes learning decisions, discourages open experimentation, and reinforces existing inequalities in access to opportunities. The value of AI-assisted learning depends not only on content quality but on how well tools support verification, credentialing, and integration with situated peer knowledge. The design recommendations we offer focus on reducing verification burden, supporting market-aware learning, and helping workers translate AI-acquired skills into credible signals for clients.

{\bf ACKNOWLEDGMENTS.} Special thanks to the anonymous reviewers for strengthening the paper and to the freelancers who participated in the study. This work was partially supported by NSF grants 2339443 and 2403252. We used Claude and Grammarly to refine language and clarity under full author oversight; all study design, analysis, literature review, and writing were conducted and verified by the authors.

\bibliographystyle{ACM-Reference-Format}
\bibliography{base}

@article{pallais2014inefficient,
  title={Inefficient hiring in entry-level labor markets},
  author={Pallais, Amanda},
  journal={American Economic Review},
  volume={104},
  number={11},
  pages={3565--3599},
  year={2014},
  publisher={American Economic Association 2014 Broadway, Suite 305, Nashville, TN 37203}
}

@article{kim2026occupational,
  title={Occupational Diversity and Stratification in Platform Work: A Longitudinal Study of Online Freelancers},
  author={Kim, Pyeonghwa and Lewandowski, Taylor and Dunn, Michael and Sawyer, Steve},
  journal={arXiv preprint arXiv:2604.03517},
  year={2026}
}

@article{beier2025workplacelearning, 
title={Workplace learning and the future of work}, 
volume={18}, DOI={10.1017/iop.2024.57}, 
number={1}, 
journal={Industrial and Organizational Psychology}, 
author={Beier, Margaret E. and Saxena, Mahima and Kraiger, Kurt and Costanza, David P. and Rudolph, Cort W. and Cadiz, David M. and Petery, Gretchen A. and Fisher, Gwenith G.}, 
year={2025}, 
pages={84–109}
}

@article{hui2024short-term,
author = {Hui, Xiang and Reshef, Oren and Zhou, Luofeng},
title = {The Short-Term Effects of Generative Artificial Intelligence on Employment: Evidence from an Online Labor Market},
year = {2024},
issue_date = {November-December 2024},
publisher = {INFORMS},
address = {Linthicum, MD, USA},
volume = {35},
number = {6},
issn = {1526-5455},
url = {https://doi.org/10.1287/orsc.2023.18441},
doi = {10.1287/orsc.2023.18441},
journal = {Organization Science},
month = nov,
pages = {1977–1989},
numpages = {13}
}

@incollection{spence1978job,
  title={Job market signaling},
  author={Spence, Michael},
  booktitle={Uncertainty in economics},
  pages={281--306},
  year={1978},
  publisher={Elsevier}
}

@article{eraut2004informal,
  title={Informal learning in the workplace},
  author={Eraut*, Michael},
  journal={Studies in continuing education},
  volume={26},
  number={2},
  pages={247--273},
  year={2004},
  publisher={Taylor \& Francis}
}

@article{blau2008relation,
  title={The relation between employee organizational and professional development activities},
  author={Blau, Gary and Andersson, Lynne and Davis, Kathleen and Daymont, Tom and Hochner, Arthur and Koziara, Karen and Portwood, Jim and Holladay, Blair},
  journal={Journal of Vocational Behavior},
  volume={72},
  number={1},
  pages={123--142},
  year={2008},
  publisher={Elsevier}
}

@online{anthropiccowork,
  author       = {{Anthropic}},
  title        = {Getting Started with Cowork},
  year         = {2026},
  url          = {https://support.claude.com/en/articles/13345190-getting-started-with-cowork},
  note         = {Accessed: 2026-02-02},
  organization = {Claude Help Center}
}

@article{kasneci2023chatgpt,
  title={ChatGPT for good? On opportunities and challenges of large language models for education},
  author={Kasneci, Enkelejda and Se{\ss}ler, Kathrin and K{\"u}chemann, Stefan and Bannert, Maria and Dementieva, Daryna and Fischer, Frank and Gasser, Urs and Groh, Georg and G{\"u}nnemann, Stephan and H{\"u}llermeier, Eyke and others},
  journal={Learning and individual differences},
  volume={103},
  pages={102274},
  year={2023},
  publisher={Elsevier}
}

@article{acemoglu2022artificial,
  title={Artificial intelligence and jobs: Evidence from online vacancies},
  author={Acemoglu, Daron and Autor, David and Hazell, Jonathon and Restrepo, Pascual},
  journal={Journal of Labor Economics},
  volume={40},
  number={S1},
  pages={S293--S340},
  year={2022},
  publisher={The University of Chicago Press Chicago, IL}
}

@book{susskind2020world,
  title={A world without work: Technology, automation and how we should respond},
  author={Susskind, Daniel},
  year={2020},
  publisher={Penguin UK}
}

@article{autor2015there,
  title={Why are there still so many jobs? The history and future of workplace automation},
  author={Autor, David H},
  journal={Journal of economic perspectives},
  volume={29},
  number={3},
  pages={3--30},
  year={2015},
  publisher={American Economic Association 2014 Broadway, Suite 305, Nashville, TN 37203-2418}
}

@article{sumbal2024wind,
  title={Wind of change: how ChatGPT and big data can reshape the knowledge management paradigm?},
  author={Sumbal, Muhammad Saleem and Amber, Quratulain and Tariq, Adeel and Raziq, Muhammad Mustafa and Tsui, Eric},
  journal={Industrial Management \& Data Systems},
  volume={124},
  number={9},
  pages={2736--2757},
  year={2024},
  publisher={Emerald Publishing Limited}
}

@inproceedings{yun2025generative,
  title={Generative ai in knowledge work: Design implications for data navigation and decision-making},
  author={Yun, Bhada and Feng, Dana and Chen, Ace S and Nikzad, Afshin and Salehi, Niloufar},
  booktitle={Proceedings of the 2025 CHI Conference on Human Factors in Computing Systems},
  pages={1--19},
  year={2025}
}

@article{toxtli2021quantifying,
  title={Quantifying the invisible labor in crowd work},
  author={Toxtli, Carlos and Suri, Siddharth and Savage, Saiph},
  journal={Proceedings of the ACM on human-computer interaction},
  volume={5},
  number={CSCW2},
  pages={1--26},
  year={2021},
  publisher={ACM New York, NY, USA}
}

@article{anderson1995cognitive,
  title={Cognitive tutors: Lessons learned},
  author={Anderson, John R and Corbett, Albert T and Koedinger, Kenneth R and Pelletier, Ray},
  journal={The journal of the learning sciences},
  volume={4},
  number={2},
  pages={167--207},
  year={1995},
  publisher={Taylor \& Francis}
}

@article{kochmar2022automated,
  title={Automated data-driven generation of personalized pedagogical interventions in intelligent tutoring systems},
  author={Kochmar, Ekaterina and Vu, Dung Do and Belfer, Robert and Gupta, Varun and Serban, Iulian Vlad and Pineau, Joelle},
  journal={International Journal of Artificial Intelligence in Education},
  volume={32},
  number={2},
  pages={323--349},
  year={2022},
  publisher={Springer}
}

@incollection{brusilovsky2007user,
  title={User models for adaptive hypermedia and adaptive educational systems},
  author={Brusilovsky, Peter and Mill{\'a}n, Eva},
  booktitle={The adaptive web: methods and strategies of web personalization},
  pages={3--53},
  year={2007},
  publisher={Springer}
}

@inproceedings{kazemitabaar2023studying,
  title={Studying the effect of AI code generators on supporting novice learners in introductory programming},
  author={Kazemitabaar, Majeed and Chow, Justin and Ma, Carl Ka To and Ericson, Barbara J and Weintrop, David and Grossman, Tovi},
  booktitle={Proceedings of the 2023 CHI conference on human factors in computing systems},
  pages={1--23},
  year={2023}
}

@inproceedings{lytvyn2025human,
  title={Human-AI Interaction in Language Acquisition: Evaluating LLM as a Language Partner},
  author={Lytvyn, Oleksandr},
  booktitle={Proceedings of the MEi: CogSci Conference},
  volume={19},
  number={1},
  year={2025}
}

@inproceedings{BrushItOff,
author = {Ma, Ning F. and Rivera, Veronica A. and Yao, Zheng and Yoon, Dongwook},
title = {“Brush it Off”: How Women Workers Manage and Cope with Bias and Harassment in Gender-agnostic Gig Platforms},
year = {2022},
isbn = {9781450391573},
publisher = {Association for Computing Machinery},
address = {New York, NY, USA},
url = {https://doi.org/10.1145/3491102.3517524},
doi = {10.1145/3491102.3517524},
booktitle = {Proceedings of the 2022 CHI Conference on Human Factors in Computing Systems},
articleno = {397},
numpages = {13},
location = {New Orleans, LA, USA},
series = {CHI '22}
}

@inproceedings{wang2025learnmate,
  title={LearnMate: Enhancing Online Education with LLM-Powered Personalized Learning Plans and Support},
  author={Wang, Xinyu Jessica and Lee, Christine P and Mutlu, Bilge},
  booktitle={Proceedings of the Extended Abstracts of the CHI Conference on Human Factors in Computing Systems},
  pages={1--10},
  year={2025}
}

@inproceedings{LLM-BasedPythonTutorCHI25,
author = {Shochcho, Muhtasim Ibteda and Rahman, Mohammad Ashfaq Ur and Rohan, Shadman and Islam, Ashraful and Heickal, Hasnain and Rahman, AKM Mahbubur and Amin, M. Ashraful and Ali, Amin Ahsan},
title = {Improving User Engagement and Learning Outcomes in LLM-Based Python Tutor: A Study of PACE},
year = {2025},
isbn = {9798400713958},
publisher = {Association for Computing Machinery},
address = {New York, NY, USA},
url = {https://doi.org/10.1145/3706599.3720240},
doi = {10.1145/3706599.3720240},
booktitle = {Proceedings of the Extended Abstracts of the CHI Conference on Human Factors in Computing Systems},
articleno = {337},
numpages = {12},
location = {
},
series = {CHI EA '25}
}

@inproceedings{ScaffoldingAlgorithmicusingLLM,
author = {Ma, Shuai and Wang, Junling and Zhang, Yuanhao and Ma, Xiaojuan and Wang, April Yi},
title = {DBox: Scaffolding Algorithmic Programming Learning through Learner-LLM Co-Decomposition},
year = {2025},
isbn = {9798400713941},
publisher = {Association for Computing Machinery},
address = {New York, NY, USA},
url = {https://doi.org/10.1145/3706598.3713748},
doi = {10.1145/3706598.3713748},
booktitle = {Proceedings of the 2025 CHI Conference on Human Factors in Computing Systems},
articleno = {585},
numpages = {20},
location = {
},
series = {CHI '25}
}

@inproceedings{LLMCompanionsCHI24,
author = {Chen, John and Lu, Xi and Du, Yuzhou and Rejtig, Michael and Bagley, Ruth and Horn, Mike and Wilensky, Uri},
title = {Learning Agent-based Modeling with LLM Companions: Experiences of Novices and Experts Using ChatGPT \& NetLogo Chat},
year = {2024},
isbn = {9798400703300},
publisher = {Association for Computing Machinery},
address = {New York, NY, USA},
url = {https://doi.org/10.1145/3613904.3642377},
doi = {10.1145/3613904.3642377},
booktitle = {Proceedings of the 2024 CHI Conference on Human Factors in Computing Systems},
articleno = {141},
numpages = {18},
location = {Honolulu, HI, USA},
series = {CHI '24}
}

@inproceedings{zamfirescu2023johnny,
  title={Why Johnny can’t prompt: how non-AI experts try (and fail) to design LLM prompts},
  author={Zamfirescu-Pereira, J Diego and Wong, Richmond Y and Hartmann, Bjoern and Yang, Qian},
  booktitle={Proceedings of the 2023 CHI conference on human factors in computing systems},
  pages={1--21},
  year={2023}
}

@inproceedings{ChallangeinLLMforlearning,
author = {Ravi, Prerna and Masla, John and Kakoti, Gisella and Lin, Grace C. and Anderson, Emma and Taylor, Matt and Ostrowski, Anastasia K. and Breazeal, Cynthia and Klopfer, Eric and Abelson, Hal},
title = {Co-designing Large Language Model Tools for Project-Based Learning with K12 Educators},
year = {2025},
isbn = {9798400713941},
publisher = {Association for Computing Machinery},
address = {New York, NY, USA},
url = {https://doi.org/10.1145/3706598.3713971},
doi = {10.1145/3706598.3713971},
booktitle = {Proceedings of the 2025 CHI Conference on Human Factors in Computing Systems},
articleno = {138},
numpages = {25},
location = {
},
series = {CHI '25}
}

@inproceedings{VerificationofLLMLabels,
author = {Wang, Xinru and Kim, Hannah and Rahman, Sajjadur and Mitra, Kushan and Miao, Zhengjie},
title = {Human-LLM Collaborative Annotation Through Effective Verification of LLM Labels},
year = {2024},
isbn = {9798400703300},
publisher = {Association for Computing Machinery},
address = {New York, NY, USA},
url = {https://doi.org/10.1145/3613904.3641960},
doi = {10.1145/3613904.3641960},
booktitle = {Proceedings of the 2024 CHI Conference on Human Factors in Computing Systems},
articleno = {303},
numpages = {21},
location = {Honolulu, HI, USA},
series = {CHI '24}
}

@inproceedings{danry2025deceptive,
  title={Deceptive explanations by large language models lead people to change their beliefs about misinformation more often than honest explanations},
  author={Danry, Valdemar and Pataranutaporn, Pat and Groh, Matthew and Epstein, Ziv},
  booktitle={Proceedings of the 2025 CHI Conference on Human Factors in Computing Systems},
  pages={1--31},
  year={2025}
}

@inproceedings{wagman2025generative,
  title={Generative AI Uses and Risks for Knowledge Workers in a Science Organization},
  author={Wagman, Kelly B and Dearing, Matthew T and Chetty, Marshini},
  booktitle={Proceedings of the 2025 CHI Conference on Human Factors in Computing Systems},
  pages={1--17},
  year={2025}
}

@article{lang2023workforce,
  title={Workforce upskilling: can universities meet the challenges of lifelong learning?},
  author={Lang, Josephine},
  journal={The International Journal of Information and Learning Technology},
  volume={40},
  number={5},
  pages={388--400},
  year={2023},
  publisher={Emerald Publishing Limited}
}

@article{roy2020future,
  title={Future of gig economy: opportunities and challenges},
  author={Roy, Gobinda and Shrivastava, Avinash K},
  journal={Imi Konnect},
  volume={9},
  number={1},
  pages={14--27},
  year={2020}
}

@article{wood2019good,
  title={Good gig, bad gig: autonomy and algorithmic control in the global gig economy},
  author={Wood, Alex J and Graham, Mark and Lehdonvirta, Vili and Hjorth, Isis},
  journal={Work, employment and society},
  volume={33},
  number={1},
  pages={56--75},
  year={2019},
  publisher={Sage Publications Sage UK: London, England}
}

@article{gagne2022understanding,
  title={Understanding and shaping the future of work with self-determination theory},
  author={Gagn{\'e}, Maryl{\`e}ne and Parker, Sharon K and Griffin, Mark A and Dunlop, Patrick D and Knight, Caroline and Klonek, Florian E and Parent-Rocheleau, Xavier},
  journal={Nature Reviews Psychology},
  volume={1},
  number={7},
  pages={378--392},
  year={2022},
  publisher={Nature Publishing Group US New York}
}

@article{mohlmann2021algorithmic,
  title={Algorithmic management of work on online labor platforms: When matching meets control.},
  author={M{\"o}hlmann, Mareike and Zalmanson, Lior and Henfridsson, Ola and Gregory, Robert Wayne},
  journal={MIS quarterly},
  volume={45},
  number={4},
  year={2021}
}

@techreport{asker2016competitive,
  title={The competitive effects of information sharing},
  author={Asker, John and Fershtman, Chaim and Jeon, Jihye and Pakes, Ariel},
  year={2016},
  institution={National Bureau of Economic Research}
}

@inproceedings{flores2025impact,
  title={The Impact of Generative AI Coding Assistants on Developers Who Are Visually Impaired},
  author={Flores-Saviaga, Claudia and Hanrahan, Benjamin V and Imteyaz, Kashif and Clarke, Steven and Savage, Saiph},
  booktitle={Proceedings of the 2025 CHI Conference on Human Factors in Computing Systems},
  pages={1--17},
  year={2025}
}

@article{chiang2018crowd,
  title={Crowd coach: Peer coaching for crowd workers' skill growth},
  author={Chiang, Chun-Wei and Kasunic, Anna and Savage, Saiph},
  journal={Proceedings of the ACM on Human-Computer Interaction},
  volume={2},
  number={CSCW},
  pages={1--17},
  year={2018},
  publisher={ACM New York, NY, USA}
}

@inproceedings{imteyaz2024human,
  title={Human Computation, Equitable, and Innovative Future of Work AI Tools},
  author={Imteyaz, Kashif and Flores Saviaga, Claudia and Savage, Saiph},
  booktitle={Proceedings of the Twelfth AAAI Conference on Human Computation and Crowdsourcing},
  volume={12},
  number={1},
  pages={155--156},
  year={2024},
  publisher={AAAI Press},
  doi={10.1609/hcomp.v12i1.31611}
}

@article{verenikina2008scaffolding,
  title={Scaffolding and learning: Its role in nurturing new learners},
  author={Verenikina, Irina},
  year={2008},
  publisher={University of Wollongong}
}

@article{seetharaman2021delivery,
  title={Delivery work and the experience of social isolation},
  author={Seetharaman, Bhavani and Pal, Joyojeet and Hui, Julie},
  journal={Proceedings of the ACM on Human-Computer Interaction},
  volume={5},
  number={CSCW1},
  pages={1--17},
  year={2021},
  publisher={ACM New York, NY, USA}
}

@incollection{pea2018social,
  title={The social and technological dimensions of scaffolding and related theoretical concepts for learning, education, and human activity},
  author={Pea, Roy D},
  booktitle={Scaffolding},
  pages={423--451},
  year={2018},
  publisher={Psychology Press}
}

@inproceedings{qiu2025self,
  title={Self-Reflective Crowds: Surfacing Wisdom through Emergent Scaffolding},
  author={Qiu, Ruo Ning and Vadaparty, Annapurna and Vintha, Suma and Dow, Steven P},
  booktitle={Proceedings of the ACM Collective Intelligence Conference},
  pages={169--187},
  year={2025}
}

@inproceedings{pachera2026co,
  title={Co-Data: Cultivating Effective Human-LLM Collaboration for Collaborative Data Processing},
  author={Pachera, Amedeo and Mauri, Andrea and Imteyaz, Kashif and Yang, Jie and Umuhoza, Eric and Bonifati, Angela and Lahav, Michal and Goyal, Nitesh},
  booktitle={Proceedings of the Extended Abstracts of the 2026 CHI Conference on Human Factors in Computing Systems},
  pages={1--7},
  year={2026}
}

@article{imteyaz2024gigsense,
  title={GigSense: An LLM-Infused Tool forWorkers' Collective Intelligence},
  author={Imteyaz, Kashif and Flores-Saviaga, Claudia and Savage, Saiph},
  journal={arXiv preprint arXiv:2405.02528},
  year={2024}
}

@inproceedings{imteyaz2026co,
  title={Co-Designing Collaborative Generative AI Tools for Freelancers},
  author={Imteyaz, Kashif and Muller, Michael and Flores-Saviaga, Claudia and Savage, Saiph},
  booktitle={Proceedings of the 2026 CHI Conference on Human Factors in Computing Systems},
  pages={1--21},
  year={2026}
}

@article{tamkin2021understanding,
  title={Understanding the capabilities, limitations, and societal impact of large language models},
  author={Tamkin, Alex and Brundage, Miles and Clark, Jack and Ganguli, Deep},
  journal={arXiv preprint arXiv:2102.02503},
  year={2021}
}

@misc{statista_freelancers,
  author       = {Statista},
  title        = {Amount of people freelancing U.S. 2023},
  year         = {2023},
  url          = {https://www.statista.com/statistics/685468/amount-of-people-freelancing-us/},
  note         = {Accessed: 2025-04-01}
}

@article{wood2018workers,
  title={Workers of the Internet unite? Online freelancer organisation among remote gig economy workers in six Asian and African countries},
  author={Wood, Alex J and Lehdonvirta, Vili and Graham, Mark},
  journal={New Technology, Work and Employment},
  volume={33},
  number={2},
  pages={95--112},
  year={2018},
  publisher={Wiley Online Library}
}

@article{huang2024design,
  title={Design Tensions in Online Freelancing Platforms: Using Speculative Participatory Design to Support Freelancers' Relationships with Clients},
  author={Huang, Jessica and Ma, Ning F and Rivera, Veronica A and Somani, Tabreek and Lee, Patrick Yung Kang and Mcgrenere, Joanna and Yoon, Dongwook},
  journal={Proceedings of the ACM on Human-Computer Interaction},
  volume={8},
  number={CSCW1},
  pages={1--28},
  year={2024},
  publisher={ACM New York, NY, USA}
}

@article{pleskac2021ecology,
  title={The ecology of competition: A theory of risk--reward environments in adaptive decision making.},
  author={Pleskac, Timothy J and Conradt, Larissa and Leuker, Christina and Hertwig, Ralph},
  journal={Psychological Review},
  volume={128},
  number={2},
  pages={315},
  year={2021},
  publisher={American Psychological Association}
}

@article{cotter2020algorithmic,
  title={Algorithmic knowledge gaps: A new horizon of (digital) inequality},
  author={Cotter, Kelley and Reisdorf, Bianca C},
  journal={International Journal of Communication},
  volume={14},
  pages={21},
  year={2020}
}

@article{blaising2022managing,
  title={Managing the transition to online freelance platforms: self-directed socialization},
  author={Blaising, Allie and Dabbish, Laura},
  journal={Proceedings of the ACM on Human-Computer Interaction},
  volume={6},
  number={CSCW2},
  pages={1--26},
  year={2022},
  publisher={ACM New York, NY, USA}
}

@article{alvarez2023understanding,
  title={Understanding platform mediated work-life: a diary study with gig economy freelancers},
  author={Alvarez de la Vega, Juan Carlos and Cecchinato, Marta E and Rooksby, John and Newbold, Joseph},
  journal={Proceedings of the ACM on Human-Computer Interaction},
  volume={7},
  number={CSCW1},
  pages={1--32},
  year={2023},
  publisher={ACM New York, NY, USA}
}

@article{prasad2024towards,
  title={Towards adoption of generative AI in organizational settings},
  author={Prasad Agrawal, Kalyan},
  journal={Journal of Computer Information Systems},
  volume={64},
  number={5},
  pages={636--651},
  year={2024},
  publisher={Taylor \& Francis}
}

@misc{upwork_freelancing_2023,
  author       = {Upwork},
  title        = {Upwork Study Finds 64 Million Americans Freelanced in 2023, Adding \$1.27 Trillion to U.S. Economy},
  year         = {2023},
  url          = {https://investors.upwork.com/news-releases/news-release-details/upwork-study-finds-64-million-americans-freelanced-2023-adding},
  note         = {Accessed: 2025-04-01}
}

@incollection{haraway2013situated,
  title={Situated knowledges: The science question in feminism and the privilege of partial perspective 1},
  author={Haraway, Donna},
  booktitle={Women, science, and technology},
  pages={455--472},
  year={2013},
  publisher={Routledge}
}

@article{neureiter2017exploring,
  title={Exploring User Requirements for Online Cooperation through Social Capital Theory.},
  author={Neureiter, Katja and Bruckenberger, Ulrike and Krischkowsky, Alina and Tscheligi, Manfred},
  journal={IxD\&A},
  volume={34},
  pages={187--206},
  year={2017}
}

@article{fan2020crowdco,
  title={Crowdco-op: Sharing risks and rewards in crowdsourcing},
  author={Fan, Shaoyang and Gadiraju, Ujwal and Checco, Alessandro and Demartini, Gianluca},
  journal={Proceedings of the ACM on Human-Computer Interaction},
  volume={4},
  number={CSCW2},
  pages={1--24},
  year={2020},
  publisher={ACM New York, NY, USA}
}

@article{CooperationRied,
author = {Fulker, Zachary and Riedl, Christoph},
title = {Cooperation in the Gig Economy: Insights from Upwork Freelancers},
year = {2024},
issue_date = {April 2024},
publisher = {Association for Computing Machinery},
address = {New York, NY, USA},
volume = {8},
number = {CSCW1},
url = {https://doi.org/10.1145/3637314},
doi = {10.1145/3637314},
journal = {Proc. ACM Hum.-Comput. Interact.},
month = apr,
articleno = {37},
numpages = {20}
}

@article{wood1976role,
  title={The role of tutoring in problem solving},
  author={Wood, David and Bruner, Jerome S and Ross, Gail},
  journal={Journal of child psychology and psychiatry},
  volume={17},
  number={2},
  pages={89--100},
  year={1976},
  publisher={Blackwell Publishing Ltd Oxford, UK}
}

@article{stein2001textbook,
  title={Textbook evaluation and adoption},
  author={Stein, Marcy and Stuen, Carol and Carnine, Douglas and Long, Roger M},
  journal={Reading \& Writing Quarterly},
  volume={17},
  number={1},
  pages={5--23},
  year={2001},
  publisher={Informa UK Limited}
}

@article{conole2005learning,
  title={A learning design toolkit to create pedagogically effective learning activities},
  author={Conole, Gr{\'a}inne and Fill, Karen},
  journal={Journal of Interactive Media in Education},
  number={1},
  year={2005}
}

@incollection{reiser2018scaffolding,
  title={Scaffolding complex learning: The mechanisms of structuring and problematizing student work},
  author={Reiser, Brian J},
  booktitle={Scaffolding},
  pages={273--304},
  year={2018},
  publisher={Psychology Press}
}

@inproceedings{quintana2004ideakeeper,
  title={IdeaKeeper Notepads: Scaffolding digital library information analysis in online inquiry},
  author={Quintana, Chris and Zhang, Meilan},
  booktitle={CHI'04 Extended Abstracts on Human Factors in Computing Systems},
  pages={1329--1332},
  year={2004}
}

@article{riedl2025potential,
  title={The potential and challenges of AI for collective intelligence},
  author={Riedl, Christoph and De Cremer, David and Lucarelli, Gina and Antoine-Souklaye, Erika and Bullock, Seth and Ajmeri, Nirav and Batty, Mike and Black, Michaela and Cartlidge, John and Challen, Robert and others},
  journal={Collective Intelligence},
  volume={4},
  number={1},
  pages={26339137241308821},
  year={2025},
  publisher={Sage Publications Sage UK: London, England}
}

@article{popiel2017boundaryless,
  title={“Boundaryless” in the creative economy: assessing freelancing on Upwork},
  author={Popiel, Pawel},
  journal={Critical Studies in Media Communication},
  volume={34},
  number={3},
  pages={220--233},
  year={2017},
  publisher={Taylor \& Francis}
}

@inproceedings{YouApp,
author = {Stein, Jake M L and Vizgirda, Vidminas and Van Kleek, Max and Binns, Reuben and Zhao, Jun and Zhao, Rui and Goel, Naman and Chalhoub, George and Albayaydh, Wael S and Shadbolt, Nigel},
title = {‘You are you and the app. There’s nobody else.’: Building Worker-Designed Data Institutions within Platform Hegemony},
year = {2023},
isbn = {9781450394215},
publisher = {Association for Computing Machinery},
address = {New York, NY, USA},
url = {https://doi.org/10.1145/3544548.3581114},
doi = {10.1145/3544548.3581114},
booktitle = {Proceedings of the 2023 CHI Conference on Human Factors in Computing Systems},
articleno = {281},
numpages = {26},
location = {Hamburg, Germany},
series = {CHI '23}
}

@article{braun2019reflecting,
  title={Reflecting on reflexive thematic analysis},
  author={Braun, Virginia and Clarke, Victoria},
  journal={Qualitative research in sport, exercise and health},
  volume={11},
  number={4},
  pages={589--597},
  year={2019},
  publisher={Taylor \& Francis}
}

@article{simkute2025ironies,
  title={Ironies of generative AI: understanding and mitigating productivity loss in Human-AI interaction},
  author={Simkute, Auste and Tankelevitch, Lev and Kewenig, Viktor and Scott, Ava Elizabeth and Sellen, Abigail and Rintel, Sean},
  journal={International Journal of Human--Computer Interaction},
  volume={41},
  number={5},
  pages={2898--2919},
  year={2025},
  publisher={Taylor \& Francis}
}

@article{shao2025future,
  title={Future of Work with AI Agents: Auditing Automation and Augmentation Potential across the US Workforce},
  author={Shao, Yijia and Zope, Humishka and Jiang, Yucheng and Pei, Jiaxin and Nguyen, David and Brynjolfsson, Erik and Yang, Diyi},
  journal={arXiv preprint arXiv:2506.06576},
  year={2025}
}

@article{Claudepaper,
  title={Which economic tasks are performed with ai? evidence from millions of claude conversations},
  author={Handa, Kunal and Tamkin, Alex and McCain, Miles and Huang, Saffron and Durmus, Esin and Heck, Sarah and Mueller, Jared and Hong, Jerry and Ritchie, Stuart and Belonax, Tim and others},
  journal={arXiv preprint arXiv:2503.04761},
  year={2025}
}

@inproceedings{do2024designing,
  title={Designing gig worker sousveillance tools},
  author={Do, Kimberly and De Los Santos, Maya and Muller, Michael and Savage, Saiph},
  booktitle={Proceedings of the 2024 CHI Conference on Human Factors in Computing Systems},
  pages={1--19},
  year={2024}
}

@inproceedings{woodruff2024knowledge,
  title={How knowledge workers think generative ai will (not) transform their industries},
  author={Woodruff, Allison and Shelby, Renee and Kelley, Patrick Gage and Rousso-Schindler, Steven and Smith-Loud, Jamila and Wilcox, Lauren},
  booktitle={Proceedings of the 2024 CHI Conference on Human Factors in Computing Systems},
  pages={1--26},
  year={2024}
}

@inproceedings{munoz2022new,
  title={New futures of work or continued marginalization? The rise of online freelance work and digital platforms},
  author={Munoz, Isabel and Sawyer, Steve and Dunn, Michael},
  booktitle={Proceedings of the 1st Annual Meeting of the Symposium on Human-Computer Interaction for Work},
  pages={1--7},
  year={2022}
}

@inproceedings{adnin2025examining,
  title={Examining Student and Teacher Perspectives on Undisclosed Use of Generative AI in Academic Work},
  author={Adnin, Rudaiba and Pandkar, Atharva and Yao, Bingsheng and Wang, Dakuo and Das, Maitraye},
  booktitle={Proceedings of the 2025 CHI Conference on Human Factors in Computing Systems},
  pages={1--17},
  year={2025}
}

@inproceedings{venkatesh2023measure,
  title={Measure of well-being of freelancers in IT sector},
  author={Venkatesh, PSKP and Selvakumar, V and Ramu, M and Manikandan, M and Senthilnathan, CR},
  booktitle={2023 Intelligent Computing and Control for Engineering and Business Systems (ICCEBS)},
  pages={1--4},
  year={2023},
  organization={IEEE}
}

@inproceedings{hsieh2025gig2gether,
  title={Gig2Gether: Datasharing to Empower, Unify and Demystify Gig Work},
  author={Hsieh, Jane and Zhang, Angie and Surati, Sajel and Xie, Sijia and Ayala, Yeshua and Sathiya, Nithila and Kuo, Tzu-Sheng and Lee, Min Kyung and Zhu, Haiyi},
  booktitle={Proceedings of the 2025 CHI Conference on Human Factors in Computing Systems},
  pages={1--25},
  year={2025}
}

@inproceedings{wei2024person,
  title={Person-Career Fit: An Exploration of Self-Employed Freelancers},
  author={Wei, Qi and Whiley, Lilith and Boyd, Caroline},
  booktitle={Academy of Management Proceedings},
  volume={2024},
  number={1},
  pages={13093},
  year={2024},
  organization={Academy of Management Valhalla, NY 10595}
}

@inproceedings{zhang2022algorithmic,
  title={Algorithmic management reimagined for workers and by workers: Centering worker well-being in gig work},
  author={Zhang, Angie and Boltz, Alexander and Wang, Chun Wei and Lee, Min Kyung},
  booktitle={Proceedings of the 2022 CHI conference on human factors in computing systems},
  pages={1--20},
  year={2022}
}

@inproceedings{calacci2022organizing,
  title={Organizing in the end of employment: information sharing, data stewardship, and digital workerism},
  author={Calacci, Dan},
  booktitle={Proceedings of the 1st Annual Meeting of the Symposium on Human-Computer Interaction for Work},
  pages={1--9},
  year={2022}
}

@inproceedings{zhang2024data,
  title={Data probes as boundary objects for technology policy design: demystifying technology for policymakers and aligning stakeholder objectives in rideshare gig work},
  author={Zhang, Angie and Rana, Rocita and Boltz, Alexander and Dubal, Veena and Lee, Min Kyung},
  booktitle={Proceedings of the 2024 CHI Conference on Human Factors in Computing Systems},
  pages={1--21},
  year={2024}
}

@inproceedings{hsieh2023co,
  title={Co-designing alternatives for the future of gig worker well-being: Navigating multi-stakeholder incentives and preferences},
  author={Hsieh, Jane and Karger, Miranda and Zagal, Lucas and Zhu, Haiyi},
  booktitle={Proceedings of the 2023 ACM Designing Interactive Systems Conference},
  pages={664--687},
  year={2023}
}

@article{nagaraj2025rideshare,
  title={Rideshare transparency: Translating gig worker insights on ai platform design to policy},
  author={Nagaraj Rao, Varun and Dalal, Samantha and Agarwal, Eesha and Calacci, Dana and Monroy-Hern{\'a}ndez, Andr{\'e}s},
  journal={Proceedings of the ACM on Human-Computer Interaction},
  volume={9},
  number={2},
  pages={1--49},
  year={2025},
  publisher={ACM New York, NY, USA}
}

@inproceedings{knight2024impact,
  title={The impact of AI technology on the productivity of gig economy workers},
  author={Knight, Benjamin and Mitrofanov, Dmitry and Netessine, Serguei},
  booktitle={Proceedings of the 25th ACM Conference on Economics and Computation},
  pages={833--833},
  year={2024}
}

@article{blaising2021making,
  title={Making it work, or not: A longitudinal study of career trajectories among online freelancers},
  author={Blaising, Allie and Kotturi, Yasmine and Kulkarni, Chinmay and Dabbish, Laura},
  journal={Proceedings of the ACM on Human-Computer Interaction},
  volume={4},
  number={CSCW3},
  pages={1--29},
  year={2021},
  publisher={ACM New York, NY, USA}
}

@inproceedings{polimetla2026paradigm,
  title={A Paradigm for Creative Ownership},
  author={Polimetla, Tejaswi and Gero, Katy Ilonka and Glassman, Elena L},
  booktitle={Proceedings of the 2026 CHI Conference on Human Factors in Computing Systems},
  pages={1--16},
  year={2026}
}

@inproceedings{he2025contributions,
  title={Which contributions deserve credit? Perceptions of attribution in human-AI co-creation},
  author={He, Jessica and Houde, Stephanie and Weisz, Justin D},
  booktitle={Proceedings of the 2025 CHI conference on human factors in computing systems},
  pages={1--18},
  year={2025}
}

@article{wilkins2022gigified,
  title={Gigified knowledge work: understanding knowledge gaps when knowledge work and on-demand work intersect},
  author={Wilkins, Denise J and Hulikal Muralidhar, Srihari and Meijer, Max and Lascau, Laura and Lindley, Si{\^a}n},
  journal={Proceedings of the ACM on human-computer interaction},
  volume={6},
  number={CSCW1},
  pages={1--27},
  year={2022},
  publisher={ACM New York, NY, USA}
}

@article{shen2026ai,
  title={How AI Impacts Skill Formation},
  author={Shen, Judy Hanwen and Tamkin, Alex},
  journal={arXiv preprint arXiv:2601.20245},
  year={2026}
}

@article{brynjolfsson2025generative,
  title={Generative AI at work},
  author={Brynjolfsson, Erik and Li, Danielle and Raymond, Lindsey},
  journal={The Quarterly Journal of Economics},
  volume={140},
  number={2},
  pages={889--942},
  year={2025},
  publisher={Oxford University Press}
}

@inproceedings{zhang2025knowledge,
  title={Knowledge Workers' Perspectives on AI Training for Responsible AI Use},
  author={Zhang, Angie and Lee, Min Kyung},
  booktitle={Proceedings of the 2025 CHI Conference on Human Factors in Computing Systems},
  pages={1--18},
  year={2025}
}

@article{systematic_gig_economy_2024,
  title={Systematic literature review on gig economy: Power dynamics, worker autonomy, and the role of social networks},
  author={Pilatti, Gustavo R and Pinheiro, Flavio L and Montini, Alessandra A},
  journal={Administrative Sciences},
  volume={14},
  number={10},
  pages={267},
  year={2024},
  publisher={MDPI}
}

@article{reskilling_upskilling_2022,
  title={Reskilling and upskilling the future-ready workforce for industry 4.0 and beyond},
  author={Li, Ling},
  journal={Information Systems Frontiers},
  volume={26},
  number={5},
  pages={1697--1712},
  year={2024},
  publisher={Springer}
}

@article{gussek2024freelancer,
  author={Gussek, Fabian},
  title={Understanding the careers of freelancers on digital labor platforms: The case of IT work},
  journal={Information Systems Journal},
  year={2024},
  publisher={Wiley},
  url={https://onlinelibrary.wiley.com/doi/10.1111/isj.12509},
  doi={10.1111/isj.12509}
}

@inproceedings{10.1145/3613905.3636293,
author = {Prpa, Mirjana and Troiano, Giovanni Maria and Wood, Matthew and Coady, Yvonne},
title = {Challenges and Opportunities of LLM-Based Synthetic Personae and Data in HCI},
year = {2024},
isbn = {9798400703317},
publisher = {Association for Computing Machinery},
address = {New York, NY, USA},
url = {https://doi.org/10.1145/3613905.3636293},
doi = {10.1145/3613905.3636293},
booktitle = {Extended Abstracts of the CHI Conference on Human Factors in Computing Systems},
articleno = {461},
numpages = {5},
location = {Honolulu, HI, USA},
series = {CHI EA '24}
}

@article{garrison1997self,
  title={Self-directed learning: Toward a comprehensive model},
  author={Garrison, D Randy},
  journal={Adult education quarterly},
  volume={48},
  number={1},
  pages={18--33},
  year={1997},
  publisher={Sage Publications Sage CA: Thousand Oaks, CA}
}

@article{jobDisplacement2024,
  title={Who is AI replacing? The impact of generative AI on online freelancing platforms},
  author={Demirci, Ozge and Hannane, Jonas and Zhu, Xinrong},
  journal={Management Science},
  volume={71},
  number={10},
  pages={8097--8108},
  year={2025},
  publisher={INFORMS}
}

@article{teutloff2025winners,
  title={Winners and losers of generative AI: Early Evidence of Shifts in Freelancer Demand},
  author={Teutloff, Ole and Einsiedler, Johanna and K{\"a}ssi, Otto and Braesemann, Fabian and Mishkin, Pamela and del Rio-Chanona, R Maria},
  journal={Journal of Economic Behavior \& Organization},
  pages={106845},
  year={2025},
  publisher={Elsevier}
}

@inproceedings{kyi2025governance,
  title={Governance of Generative AI in Creative Work: Consent, Credit, Compensation, and Beyond},
  author={Kyi, Lin and Mahuli, Amruta and Silberman, M Six and Binns, Reuben and Zhao, Jun and Biega, Asia J},
  booktitle={Proceedings of the 2025 CHI Conference on Human Factors in Computing Systems},
  pages={1--16},
  year={2025}
}

@inproceedings{10.1145/3596671.3597655, 
author = {Kim, Pyeonghwa and Sawyer, Steve}, 
title = {Many Futures of Work and Skill: Heterogeneity in Skill Building Experiences on Digital Labor Platforms}, 
year = {2023}, 
isbn = {9798400708077}, 
publisher = {Association for Computing Machinery}, 
address = {New York, NY, USA}, 
url = {https://doi.org/10.1145/3596671.3597655}, 
doi = {10.1145/3596671.3597655}, 
booktitle = {Proceedings of the 2nd Annual Meeting of the Symposium on Human-Computer Interaction for Work}, 
articleno = {11}, 
numpages = {9}, 
location = {Oldenburg, Germany}, 
series = {CHIWORK '23} 
}

@inproceedings{dolata2025moreAttention,
author = {Dolata, Mateusz and Lange, Norbert and Schwabe, Gerhard},
title = {More Attention, Transformation, Acceleration, and Exploration: Freelance Developers' Take on Hypes},
year = {2025},
isbn = {9798400713941},
publisher = {Association for Computing Machinery},
address = {New York, NY, USA},
url = {https://doi.org/10.1145/3706598.3713097},
doi = {10.1145/3706598.3713097},
booktitle = {Proceedings of the 2025 CHI Conference on Human Factors in Computing Systems},
articleno = {914},
numpages = {21},
location = {
},
series = {CHI '25}
}

@misc{peng2023impactaideveloperproductivity,
      title={The Impact of AI on Developer Productivity: Evidence from GitHub Copilot}, 
      author={Sida Peng and Eirini Kalliamvakou and Peter Cihon and Mert Demirer},
      year={2023},
      eprint={2302.06590},
      archivePrefix={arXiv},
      primaryClass={cs.SE},
      url={https://arxiv.org/abs/2302.06590}, 
}

@incollection{kitchenham2008personal,
  title={Personal opinion surveys},
  author={Kitchenham, Barbara A and Pfleeger, Shari L},
  booktitle={Guide to advanced empirical software engineering},
  pages={63--92},
  year={2008},
  publisher={Springer}
}

@book{blandford2016qualitative,
  title={Qualitative HCI research: Going behind the scenes},
  author={Blandford, Ann and Furniss, Dominic and Makri, Stephann},
  year={2016},
  publisher={Morgan \& Claypool Publishers}
}

@article{campbell2020purposive,
  title={Purposive sampling: complex or simple? Research case examples},
  author={Campbell, Steve and Greenwood, Melanie and Prior, Sarah and Shearer, Toniele and Walkem, Kerrie and Young, Sarah and Bywaters, Danielle and Walker, Kim},
  journal={Journal of research in Nursing},
  volume={25},
  number={8},
  pages={652--661},
  year={2020},
  publisher={Sage Publications Sage UK: London, England}
}

@inproceedings{Uncertainty10.1145/3290607.3312922,
author = {Blaising, Allie and Kotturi, Yasmine and Kulkarni, Chinmay},
title = {Navigating Uncertainty in the Future of Work: Information-Seeking and Critical Events Among Online Freelancers},
year = {2019},
isbn = {9781450359719},
publisher = {Association for Computing Machinery},
address = {New York, NY, USA},
url = {https://doi.org/10.1145/3290607.3312922},
doi = {10.1145/3290607.3312922},
booktitle = {Extended Abstracts of the 2019 CHI Conference on Human Factors in Computing Systems},
pages = {1–6},
numpages = {6},
location = {Glasgow, Scotland Uk},
series = {CHI EA '19}
}

@book{gray2019ghost,
title={Ghost Work: How to Stop Silicon Valley from Building a New Global Underclass},
author={Gray, Mary L. and Suri, Siddharth},
year={2019},
publisher={Houghton Mifflin Harcourt},
address={New York, NY},
isbn={978-1-328-63346-1},
url={https://marylgray.org/bio/on-demand/}
}

@article{ivankova2006using,
  title={Using mixed-methods sequential explanatory design: From theory to practice},
  author={Ivankova, Nataliya V and Creswell, John W and Stick, Sheldon L},
  journal={Field methods},
  volume={18},
  number={1},
  pages={3--20},
  year={2006},
  publisher={Sage Publications Sage CA: Thousand Oaks, CA}
}

@article{Campana2014,
  author = {Joe Campana},
  title = {Learning for work and professional development: The significance of informal learning networks of digital media industry professionals},
  journal = {International Journal of Training Research},
  volume = {12},
  number = {3},
  pages = {213-226},
  year = {2014},
  doi = {10.1080/14480220.2014.11082043}
}

@article{Trust2016,
  author = {Trust, T. and Krutka, D. G. and Carpenter, J. P.},
  title = {``Together we are better'': Professional learning networks for teachers},
  journal = {Computers \& Education},
  volume = {102},
  pages = {15--34},
  year = {2016},
  doi = {10.1016/j.compedu.2016.06.007}
}

@article{flores2020understanding,
  title={Understanding the complementary nature of paid and volunteer crowds for content creation},
  author={Flores Saviaga, Claudia and Granados, Ricardo and Savage, Liliana and Escobedo Bravo, Lizbeth Olivia and Savage, Saiph},
  year={2020}
}

@inproceedings{glassman2015learner,
  title={Learner-sourcing in an engineering class at scale},
  author={Glassman, Elena L and Terman, Christopher J and Miller, Robert C},
  booktitle={Proceedings of the Second (2015) ACM Conference on Learning@ Scale},
  pages={363--366},
  year={2015}
}

@phdthesis{kim2015learnersourcing,
  title={Learnersourcing: improving learning with collective learner activity},
  author={Kim, Juho and others},
  year={2015},
  school={Massachusetts Institute of Technology}
}

@inproceedings{bernstein2010crowd,
  title={Crowd-powered interfaces},
  author={Bernstein, Michael S},
  booktitle={Adjunct proceedings of the 23nd annual ACM symposium on User interface software and technology},
  pages={347--350},
  year={2010}
}

@article{bernstein2008personalization,
  title={Personalization via friendsourcing},
  author={Bernstein, Michael S and Tan, Desney and Smith, Greg and Czerwinski, Mary and Horvitz, Eric},
  journal={ACM Transactions on Computer-Human Interaction (TOCHI)},
  volume={17},
  number={2},
  pages={1--28},
  year={2008},
  publisher={ACM New York, NY, USA}
}

@inproceedings{rzeszotarski2014estimating,
  title={Estimating the social costs of friendsourcing},
  author={Rzeszotarski, Jeffrey M and Morris, Meredith Ringel},
  booktitle={Proceedings of the SIGCHI Conference on Human Factors in Computing Systems},
  pages={2735--2744},
  year={2014}
}

@inproceedings{gray2016crowd,
  title={The crowd is a collaborative network},
  author={Gray, Mary L and Suri, Siddharth and Ali, Syed Shoaib and Kulkarni, Deepti},
  booktitle={Proceedings of the 19th ACM conference on computer-supported cooperative work \& social computing},
  pages={134--147},
  year={2016}
}

@inproceedings{FutureofCrowd,
author = {Kittur, Aniket and Nickerson, Jeffrey V. and Bernstein, Michael and Gerber, Elizabeth and Shaw, Aaron and Zimmerman, John and Lease, Matt and Horton, John},
title = {The future of crowd work},
year = {2013},
isbn = {9781450313315},
publisher = {Association for Computing Machinery},
address = {New York, NY, USA},
url = {https://doi.org/10.1145/2441776.2441923},
doi = {10.1145/2441776.2441923},
booktitle = {Proceedings of the 2013 Conference on Computer Supported Cooperative Work},
pages = {1301–1318},
numpages = {18},
location = {San Antonio, Texas, USA},
series = {CSCW '13}
}

@article{SkillDevelopNecessity,
  author  = {Huang, Keman and Yao, Jinhui and Yin, Ming},
  year    = {2019},
  month   = {11},
  pages   = {1-23},
  title   = {Understanding the Skill Provision in Gig Economy from A Network Perspective: A Case Study of Fiverr},
  volume  = {3},
  journal = {Proceedings of the ACM on Human-Computer Interaction},
  doi     = {10.1145/3359234}
}

@article{jarrahi2021algorithmic,
  title={Algorithmic management in a work context},
  author={Jarrahi, Mohammad Hossein and Newlands, Gemma and Lee, Min Kyung and Wolf, Christine T and Kinder, Eliscia and Sutherland, Will},
  journal={Big Data \& Society},
  volume={8},
  number={2},
  pages={20539517211020332},
  year={2021},
  publisher={SAGE Publications Sage UK: London, England}
}

@inproceedings{WorkingWithMachines,
author = {Lee, Min Kyung and Kusbit, Daniel and Metsky, Evan and Dabbish, Laura},
title = {Working with Machines: The Impact of Algorithmic and Data-Driven Management on Human Workers},
year = {2015},
isbn = {9781450331456},
publisher = {Association for Computing Machinery},
address = {New York, NY, USA},
url = {https://doi.org/10.1145/2702123.2702548},
doi = {10.1145/2702123.2702548},
booktitle = {Proceedings of the 33rd Annual ACM Conference on Human Factors in Computing Systems},
pages = {1603–1612},
numpages = {10},
location = {Seoul, Republic of Korea},
series = {CHI '15}
}

@book{emotionalSupport,
  author    = {Johnson, David W. and Johnson, Roger T.},
  title     = {Learning Together and Alone: Cooperative, Competitive, and Individualistic Learning},
  edition   = {2nd},
  year      = {1987},
  publisher = {Prentice-Hall, Inc.},
  address   = {Englewood Cliffs, NJ}
}

@article{munoz2022platform,
  title={Platform-mediated markets, online freelance workers and deconstructed identities},
  author={Munoz, Isabel and Dunn, Michael and Sawyer, Steve and Michaels, Emily},
  journal={Proceedings of the ACM on Human-Computer Interaction},
  volume={6},
  number={CSCW2},
  pages={1--24},
  year={2022},
  publisher={ACM New York, NY, USA}
}

@inproceedings{reviewsAM,
author = {Holtz, David M and Scult, Liane and Suri, Siddharth},
title = {How Much Do Platform Workers Value Reviews? An Experimental Method},
year = {2022},
isbn = {9781450391573},
publisher = {Association for Computing Machinery},
address = {New York, NY, USA},
url = {https://doi.org/10.1145/3491102.3501900},
doi = {10.1145/3491102.3501900},
booktitle = {Proceedings of the 2022 CHI Conference on Human Factors in Computing Systems},
articleno = {71},
numpages = {11},
location = {New Orleans, LA, USA},
series = {CHI '22}
}

@book{knowles1980modern,
title={The modern practice of adult education: From pedagogy to andragogy},
author={Knowles, Malcolm Shepherd and others},
volume={2},
year={1980},
publisher={Cambridge Adult Education Englewood Cliffs, NJ}
}

@article{burke2020relationship,
  title={The relationship between freelance workforce intensity, business performance and job creation},
  author={Burke, Andrew and Cowling, Marc},
  journal={Small Business Economics},
  volume={55},
  number={2},
  pages={399--413},
  year={2020},
  publisher={Springer}
}

@book{creswell2017,
  title={Designing and conducting mixed methods research},
  author={Creswell, John W and Clark, Vicki L Plano},
  year={2017},
  publisher={Sage publications}
}

@article{lowe2018quantifying,
  title={Quantifying thematic saturation in qualitative data analysis},
  author={Lowe, Andrew and Norris, Anthony C and Farris, A Jane and Babbage, Duncan R},
  journal={Field methods},
  volume={30},
  number={3},
  pages={191--207},
  year={2018},
  publisher={Sage Publications Sage CA: Los Angeles, CA}
}

@incollection{Gruenewald2025Reskilling,
  author = {Gruenewald, H. and Mueller, M.},
  title = {Challenges and Opportunities in Reskilling and Upskilling},
  booktitle = {Reskilling and Upskilling in a Globalized Economy, Future of Business and Finance},
  year = {2025},
  publisher = {Springer Fachmedien Wiesbaden GmbH, part of Springer Nature},
  doi = {10.1007/978-3-658-48384-5_7},
  pages = {167--196}
}

@article{Morandini2023,
  author = {Morandini, Sofia and Fraboni, Federico and De Angelis, Marco and Puzzo, Gabriele and Giusino, Davide and Pietrantoni, Luca},
  title = {The Impact of Artificial Intelligence on Workers' Skills: Upskilling and Reskilling in Organisations},
  journal = {Informing Science: The International Journal of an Emerging Transdiscipline},
  volume = {26},
  pages = {39--68},
  year = {2023},
  doi = {10.28945/5078},
  url = {https://inform.nu/Volumes/Vol26/ISV26p039-068Morandini8895.pdf}
}

@article{morris2019,
  title={Self-directed learning: A fundamental competence in a rapidly changing world},
  author={Morris, Thomas Howard},
  journal={International Review of Education},
  volume={65},
  number={4},
  pages={633--653},
  year={2019},
  publisher={Springer}
}

@misc{upwork2025,
  title = {{The Future Workforce Index: Evolving Talent Trends in 2025 and Beyond}},
  author = {{Upwork Research Institute}},
  year = {2025},
  month = {Apr},
  day = {23},
  url = {https://www.upwork.com/research/future-workforce-index-2025},
  note = {Accessed August 18, 2025},
}

@misc{FixedTeamwork,
  title={AI Hasn't Fixed Teamwork, But It Shifted Collaborative Culture: A Longitudinal Study in a Project-Based Software Development Organization (2023-2025)},
  author={Xiao, Qing and Hu, Xinlan Emily and Whiting, Mark E and Karunakaran, Arvind and Shen, Hong and Cao, Hancheng},
  journal={arXiv preprint arXiv:2509.10956},
  year={2025}
}

@book{kolb2014experiential,
  title={Experiential learning: Experience as the source of learning and development},
  author={Kolb, David A},
  year={2014},
  publisher={FT press}
}

\clearpage
\appendix

\onecolumn
\section{Appendix}

\subsection{Survey Participant Table}

\begin{table}[!htb]
\centering
\caption{Participant Demographics: Survey Respondents}
\label{tab:demographics}
\begin{tabular}{lcc lcc}
\toprule
\textbf{Variable} & \textbf{n} & \textbf{\%} & \textbf{Variable} & \textbf{n} & \textbf{\%} \\
\midrule
\multicolumn{3}{l}{\textit{Sex}} & \multicolumn{3}{l}{\textit{Hours per Week on Freelance/Gig Work}} \\
Female & 37 & 52.1 & 0--10 hours & 18 & 25.4 \\
Male & 33 & 46.5 & 11--20 hours & 22 & 31.0 \\
Not disclosed & 1 & 1.4 & 21--30 hours & 16 & 22.5 \\
 &  &  & 31--40 hours & 15 & 21.1 \\
\midrule
\multicolumn{3}{l}{\textit{Age}} & \multicolumn{3}{l}{\textit{Experience Working as a Freelancer}} \\
18--24 & 17 & 23.9 & < 1 year & 1 & 1.4 \\
25--34 & 39 & 54.9 & 1--2 years & 23 & 32.4 \\
35--44 & 13 & 18.3 & 2--3 years & 25 & 35.2 \\
45--55 & 1 & 1.4 & 4--5 years & 13 & 18.3 \\
55+ & 1 & 1.4 & > 5 years & 9 & 12.7 \\
\midrule
\multicolumn{3}{l}{\textit{Highest Level of Education}} & & & \\
High school & 2 & 2.8 & & & \\
Some college & 8 & 11.3 & & & \\
Bachelor's & 43 & 60.6 & & & \\
Master's & 17 & 23.9 & & & \\
Doctoral & 1 & 1.4 & & & \\
\bottomrule
\end{tabular}
\end{table}

\subsection{Interview Participant Table}
\begin{table}[htbp]
\centering
\scriptsize
\setlength{\tabcolsep}{5pt}
\renewcommand{\arraystretch}{1.15}
\caption{Demographics and professional background of interview participants ($N=20$). 
All participants hold a Bachelor's degree unless otherwise noted.}
\label{tab:participant-demographics}
\begin{tabular}{@{}clllllll@{}}
\toprule
\textbf{ID} & \textbf{Age} & \textbf{Gender} & \textbf{Degree} & \textbf{Experience} & \textbf{Platforms} & \textbf{Area of Expertise} \\
\midrule
P1  & 27 & F & Doctorate   & 5+ yrs   & Upwork                   & Strategy \& Marketing \\
P2  & 29 & M & Bachelor's  & 1--2 yrs & Upwork                   & Security Engineering \\
P3  & 31 & M & Bachelor's  & 2--3 yrs & Upwork                   & Virtual Assistant \\
P4  & 28 & F & Bachelor's  & 5+ yrs   & Upwork                   & Content \& Administration \\
P5  & 23 & M & Bachelor's  & 1--2 yrs & Upwork                   & AI Engineering \\
P6  & 24 & F & Bachelor's  & 2--3 yrs & Upwork                   & UI/UX Design \\
P7  & 25 & F & Bachelor's  & 4--5 yrs & MTurk, Upwork            & Graphic Design \\
P8  & 30 & F & Bachelor's  & 4--5 yrs & Upwork                   & Virtual Assistant \\
P9  & 25 & F & Bachelor's  & 4--5 yrs & Upwork                   & Digital Marketing \\
P10 & 22 & M & Bachelor's  & 1--2 yrs & Upwork                   & Copywriting \& Email Marketing \\
P11 & 21 & M & Bachelor's  & 2--3 yrs & Upwork                   & Graphic Design \\
P12 & 22 & F & Bachelor's  & 2--3 yrs & Upwork                   & Virtual Assistant \\
P13 & 29 & M & Bachelor's  & 4--5 yrs & Upwork                   & Virtual Assistant \\
P14 & 23 & F & Bachelor's  & 1--2 yrs & Upwork, LinkedIn         & Virtual Assistant \& Graphic Design \\
P15 & 32 & M & Bachelor's  & 5+ yrs   & Freelancer.com, Upwork   & Software Engineering \\
P16 & 22 & F & Bachelor's  & 1--2 yrs & Upwork                   & Virtual Assistant \& Data Entry \\
P17 & 24 & F & Bachelor's  & 5+ yrs   & Facebook Groups, Upwork  & Digital Marketing \& Writing \\
P18 & 25 & M & Bachelor's  & 1--2 yrs & Upwork, Fiverr           & Cloud Solutions Architect \\
P19 & 27 & F & Bachelor's  & 2--3 yrs & Upwork                   & Product Management \\
P20 & 28 & F & Bachelor's  & 2--3 yrs & Upwork                   & Talent Acquisition \& Recruiting \\
\bottomrule
\end{tabular}
\end{table}

\end{document}